\documentclass{article}

\usepackage{microtype}
\usepackage{graphicx}
\usepackage{enumitem}
\usepackage{amsfonts}
\usepackage{amsmath}
\usepackage{caption}
\usepackage{subcaption}
\usepackage{booktabs} 

\usepackage{hyperref}

\DeclareMathOperator{\out}{out}
\DeclareMathOperator{\ML}{ML}
\DeclareMathOperator{\lcm}{lcm}

\usepackage[accepted]{icml2021}

\icmltitlerunning{Cyclically Equivariant Neural Decoders for Cyclic Codes}

\begin{document}

\twocolumn[
\icmltitle{Cyclically Equivariant Neural Decoders for Cyclic Codes}



\icmlsetsymbol{equal}{*}

\begin{icmlauthorlist}
\icmlauthor{Xiangyu Chen}{to}
\icmlauthor{Min Ye}{to}
\end{icmlauthorlist}

\icmlaffiliation{to}{Data Science and Information Technology Research Center, Tsinghua-Berkeley Shenzhen Institute, Tsinghua Shenzhen International Graduate School, Shenzhen, China}

\icmlcorrespondingauthor{Min Ye}{yeemmi@gmail.com}

\icmlkeywords{Neural decoder, Cyclic codes, Belief propagation, Cyclically equivariant neural network.}

\vskip 0.3in
]



\printAffiliationsAndNotice{}  

\begin{abstract}
Neural decoders were introduced as a generalization of the classic Belief Propagation (BP) decoding algorithms, where the Trellis graph in the BP algorithm is viewed as a neural network, and the weights in the Trellis graph are optimized by training the neural network. In this work, we propose a novel neural decoder for cyclic codes by exploiting their cyclically invariant property. More precisely, we impose a shift invariant structure on the weights of our neural decoder so that any cyclic shift of inputs results in the same cyclic shift of outputs. Extensive simulations with BCH codes and punctured Reed-Muller (RM) codes show that our new decoder consistently outperforms previous neural decoders when decoding cyclic codes. Finally, we propose a list decoding procedure that can significantly reduce the decoding error probability for BCH codes and punctured RM codes. For certain high-rate codes, the gap between our list decoder and the Maximum Likelihood decoder is less than $0.1$dB.
Code available at \href{https://github.com/cyclicallyneuraldecoder/CyclicallyEquivariantNeuralDecoders}{github.com/cyclicallyneuraldecoder}
\end{abstract}

\section{Introduction}

In recent years, machine learning methods have been successfully applied to the area of decoding error-correcting codes. The usage of neural networks \cite{Nachmani16,Gruber17,Cammerer17,Nachmani18,Kim18,Kim18a,Vasic18,Teng19,Buchberger20}, autoencoders \cite{Jiang19}, graph neural networks \cite{Nachmani19} and reinforcement learning \cite{Carpi19,Habib20} have demonstrated improvements over classical algorithms in decoding various families of error-correcting codes with short to moderate block length, including BCH codes, polar codes, LDPC codes and Reed-Muller (RM) codes.

In particular, one line of research pioneered by \cite{Nachmani16,Nachmani18} introduced neural decoders as a generalization of the classic Belief Propagation (BP) decoding algorithm.
More precisely, the Trellis graph in the BP algorithm is viewed as a fully connected neural network \cite{Nachmani16}, and the weights in the Trellis graph are optimized by training the neural network. Later, \cite{Nachmani18} suggested to replace the fully connected neural networks with recurrent neural networks (RNNs). More recently, the tools of graph neural networks were also introduced to this setting \cite{Nachmani19}.
While these neural BP decoders improve upon the vanilla BP decoder for a wide range of error-correcting codes, their designs rarely utilize the algebraic properties of any particular code family. On the other hand, certain algebraic properties of some code families have proven to be the key to their good performance in error correction, e.g., the recursive structure of polar codes \cite{Arikan09} and the doubly transitive property of BCH codes and RM codes \cite{Kudekar17}.

In this paper we design a novel neural decoder for an important class of codes called cyclic codes, including two extensively studied and widely applied code families---BCH codes and punctured RM codes. As suggested by their name, cyclic codes are invariant to cyclic shifts, and this property is fully exploited in the design of our new decoder. 
Inspired by the fact that the Maximum Likelihood (ML) decoder of any cyclic code is equivariant to cyclic shifts, we impose a shift invariant structure on the weights of our neural decoder so that it shares the equivariant property of the ML decoder, i.e., any cyclic shift of inputs results in the same cyclic shift of the decoding outputs.

\begin{figure*}
\centering
\begin{subfigure}{0.2\textwidth}
\centering
\includegraphics[width=\textwidth]{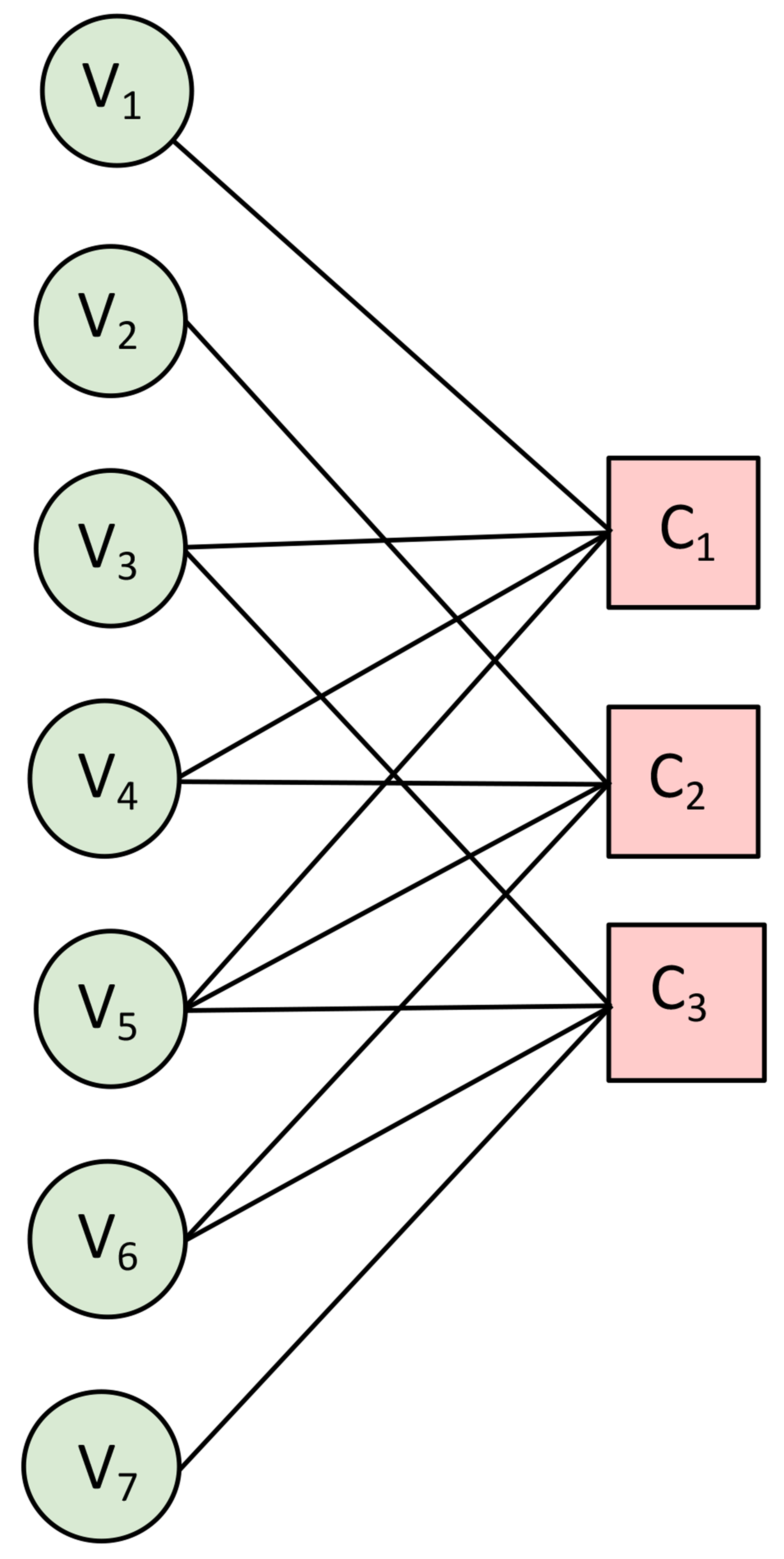} 
\caption{}
\label{fig:previous_tanner}
\end{subfigure}
\hspace*{0.6in}
\begin{subfigure}{0.65\textwidth}
\centering
\includegraphics[width=\textwidth]{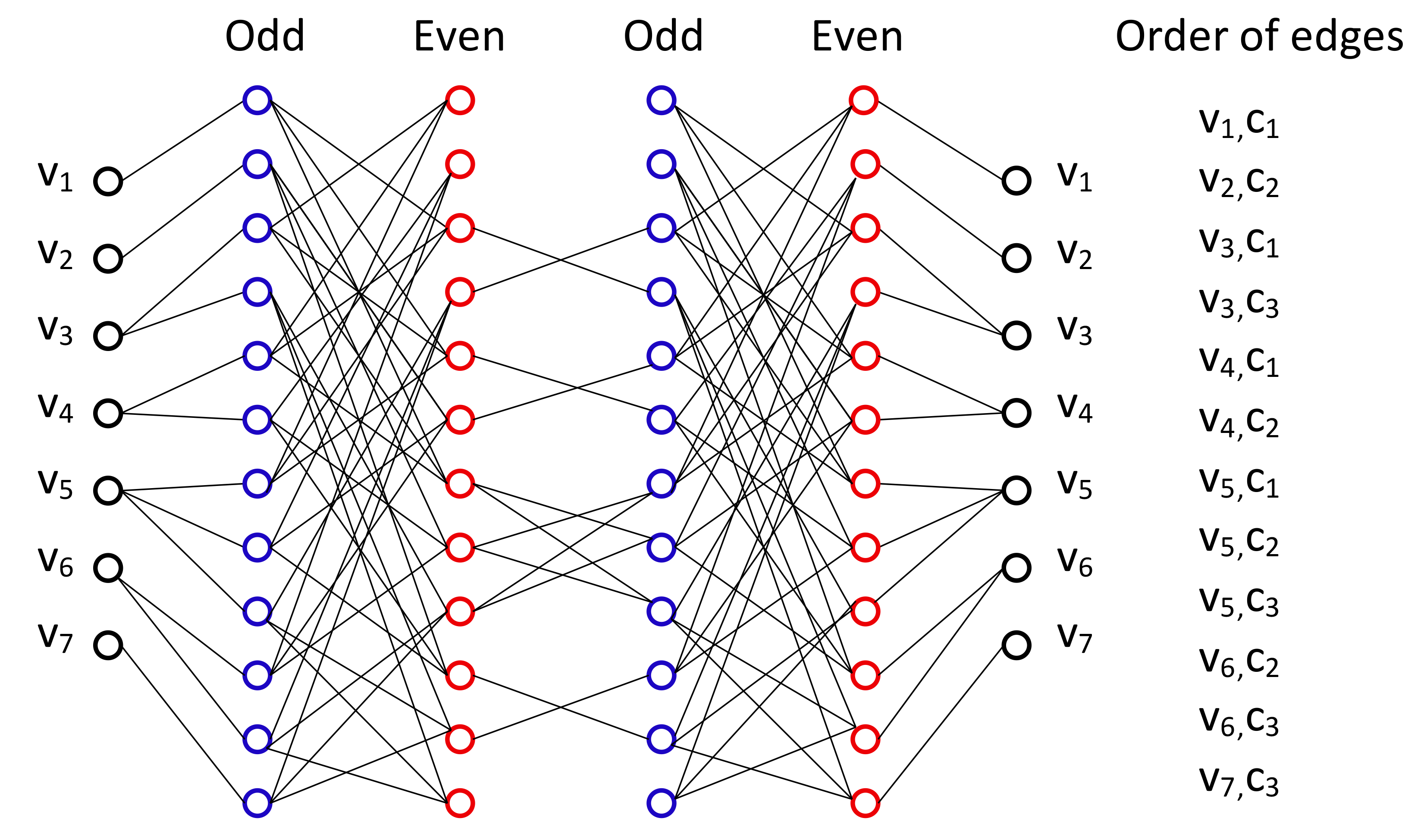} 
\caption{}
\label{fig:previous_trellis}
\end{subfigure}
\caption{(a) The Tanner graph corresponding to the parity matrix in \eqref{eq:pc74}. The variable node $v_j$ corresponds to the $j$th column of the matrix, and the check node $c_i$ corresponds to the $i$th row.
(b) The corresponding Trellis graph with two iterations. The 12 nodes in each of the 4 middle layers(columns) represent the 12 edges in the Tanner graph, whose order is listed in the rightmost column.
Node $e$ in the $s$th middle layer is connected to node $e'$ in the $(s-1)$th layer if $x^{[s-1]}(e')$ is involved in the calculation of $x^{[s]}(e)$; see \eqref{eq:classicodd}--\eqref{eq:classicout}.
}
\label{fig:previous}
\end{figure*}

We carry out extensive simulations to test the performance of our new decoder on BCH codes and punctured RM codes with short to moderate code length. Simulation results show that our decoder consistently outperforms previous neural decoders. In particular, we observe a consistent $0.7$dB improvement over the neural decoder proposed in \cite{Nachmani18} for various choices of code parameters. Our decoder also demonstrates a 0.3dB improvement over the hyper-graph-network decoder proposed in \cite{Nachmani19} with 300 times smaller training time. As a concrete example, for BCH codes with length 63 and dimension 45, it only takes 10 minutes to train our decoder while the training of the hyper-graph-network decoder \cite{Nachmani19} takes more than 2 days on the same platform.

Finally, we propose a list decoding procedure that can significantly reduce the decoding error probability for BCH codes and punctured RM codes. Our list decoding procedure exploits the rich automorphism groups of extended BCH codes and RM codes. More precisely, extended BCH codes and RM codes are obtained by adding an overall parity bit to the BCH codes and punctured RM codes, respectively, and these two extended code families are invariant to a large automorphism group of permutations. In order to make use of this property, we add a dummy symbol to the received noisy codeword and then apply the permutations in the automorphism group to the extended noisy codeword. For each permutation, our neural decoder gives us an intermediate decoding result, and the final decoding result of the list decoding procedure is given by a ML decoding among all the intermediate decoding results. Extensive simulations show that the list decoding method provides up to 3dB gain. As a final remark, we note that this list decoding method can be coupled with any decoding algorithm for BCH codes and punctured RM codes, not just our neural decoder.

To conclude this section, we summarize the {\bf main contributions} of this paper:
\begin{itemize}[nosep]
    \item We propose a novel neural decoder that is provably equivariant to cyclic shifts.
    \item Extensive simulations with BCH codes and punctured RM codes show that our neural decoder consistently improves upon \cite{Nachmani18} by $0.7$dB, and it also improves upon the hyper-graph-network decoder \cite{Nachmani19} by 0.3dB with 300 times smaller training time.
    \item We propose a list decoding procedure that provides up to 3dB gain for BCH codes and punctured RM codes.
    For certain high-rate codes, our list decoder with list size $n+1$ has almost the same performance as the ML decoder, where $n$ is the code length.
\end{itemize}

\section{Background and previous neural decoders}
A linear code with code length $n$ and code dimension $k$ is a $k$-dimensional subspace of $\mathbb{F}_2^n$, where $\mathbb{F}_2=\{0,1\}$ is the binary field. It can be defined in two equivalent ways---either by a binary generator matrix $G$ of size $k\times n$ or by a binary parity check matrix $H$ of size $(n-k)\times n$.

Each parity check matrix $H$ entails a Tanner graph, which is a bipartite graph with $n$ variable nodes labelled as $v_1,v_2,\dots,v_n$ on the left side and $n-k$ check nodes labelled as $c_1,c_2,\dots,c_{n-k}$ on the right side. An edge is connected between $v_j$ and $c_i$ in the Tanner graph if and only if $H_{ij} = 1$.
As a concrete example, a parity check matrix of the $(n=7,k=4)$ Hamming code is
\begin{equation} \label{eq:pc74}
\begin{array}{ccccccc}
1 & 0 & 1 & 1 & 1 & 0 & 0 \\
0 & 1 & 0 & 1 & 1 & 1 & 0 \\
0 & 0 & 1 & 0 & 1 & 1 & 1
\end{array} ,
\end{equation}
and the corresponding Tanner graph is given in Fig.~\ref{fig:previous_tanner}. In BP algorithms, messages propagate back and forth through the edges in the Tanner graph, and this message passing procedure is best depicted by the Trellis graph, obtained from unrolling the Tanner graph; see Fig.~\ref{fig:previous_trellis}. A Trellis graph with $t$ iterations of message passing consists of one input layer, one output layer, and $2t$ middle layers. Both the input and output layers consist of $n$ variable nodes while the nodes in each middle layer correspond to the edges in the Tanner graph.

In BP algorithms, messages propagate through the Trellis graph from left to right. The $n$ variable nodes in the input layer hold the log likelihood ratios (LLR) of the $n$ input bits: 
$$
L_j = \log \frac{\mathbb{P} (y_j|C_j=0)}{\mathbb{P} (y_j|C_j=1)} 
\quad \text{for~} j\in[n] ,
$$
where $(C_1,\dots,C_n)$ is a randomly chosen codeword, and $(y_1,\dots,y_n)$ is the channel output after transmitting $(C_1,\dots,C_n)$ through $n$ independent copies of some noisy channel. BP algorithm aims to recover the codeword from the channel output, or equivalently, from the LLRs.
Let $E$ be the set consisting of all the edges in the Tanner graph, and we use $e=(v_j,c_i)$ and $e=(c_i,v_j)$ interchangeably to denote the same edge. Let the vector $x^{[s]}=(x^{[s]}(e),e\in E)$ be the message vector held by the $s$th middle layer in the Trellis graph. Suppose that the BP algorithm has $t$ iterations. For each $s=1,2,\dots,2t$, the message vector $x^{[s]}$ is calculated recursively from $x^{[s-1]}$ and the LLR vector $(L_1,\dots,L_n)$, where the initialization $x^{[0]}$ is the all zero vector. More precisely, for odd $s$ and an edge $e=(c_i,v_j)\in E$, the message $x^{[s]}(e)$ is given by
\begin{equation} \label{eq:classicodd}
\begin{aligned}
& x^{[s]}(e)=
x^{[s]}((c_i,v_j)) \\
= & \tanh \Big( \frac{1}{2} \Big( L_j + \sum_{e'\in N(v_j)\setminus \{e\} } x^{[s-1]}(e') \Big) \Big) ,
\end{aligned}
\end{equation}
where $N(v_j)\subseteq E$ is the set of all the edges containing $v_j$ as an endpoint in the Tanner graph.
For even $s$ and an edge $e=(c_i,v_j)\in E$, the message $x^{[s]}(e)$ is given by
\begin{equation} \label{eq:classiceven}
x^{[s]}(e) 
= 2 \tanh^{-1} \Big( \prod_{e'\in N(c_i)\setminus \{e\} } x^{[s-1]}(e') \Big) ,
\end{equation}
where $N(c_i)\subseteq E$ is the set of all the edges containing $c_i$ as an endpoint in the Tanner graph.
Finally, the output of the classic BP algorithm is
\begin{equation} \label{eq:classicout}
o_j = L_j + \sum_{e\in N(v_j)} x^{[2t]}(e)
\quad \text{for~} j\in [n] .
\end{equation}

In \cite{Nachmani16,Nachmani18}, a set of learnable weights are added into the calculations of odd and output layers while the calculations of even layers, i.e., \eqref{eq:classiceven}, remain unchanged. More precisely, \eqref{eq:classicodd} and \eqref{eq:classicout} are replaced by
\begin{align}
x^{[s]}(e)=
x^{[s]} & ((c_i,v_j)) 
=  \tanh  \Big( \frac{1}{2} \Big( w^{[s]}(j,e) L_j 
\label{eq:weightedodd} \\
& + \sum_{e'\in N(v_j)\setminus \{e\} }
w^{[s]}(e', e) ~
x^{[s-1]}(e') \Big) \Big) , \nonumber \\
 \text{and~}  o_j = L_j & +  \sum_{e\in N(v_j)} 
w^{\out}(e, j) ~
x^{[2t]}(e) , \label{eq:weightedout}
\end{align}
respectively.

\begin{figure*}
\centering
\begin{subfigure}{0.49\textwidth}
\centering
\includegraphics[width=\textwidth]{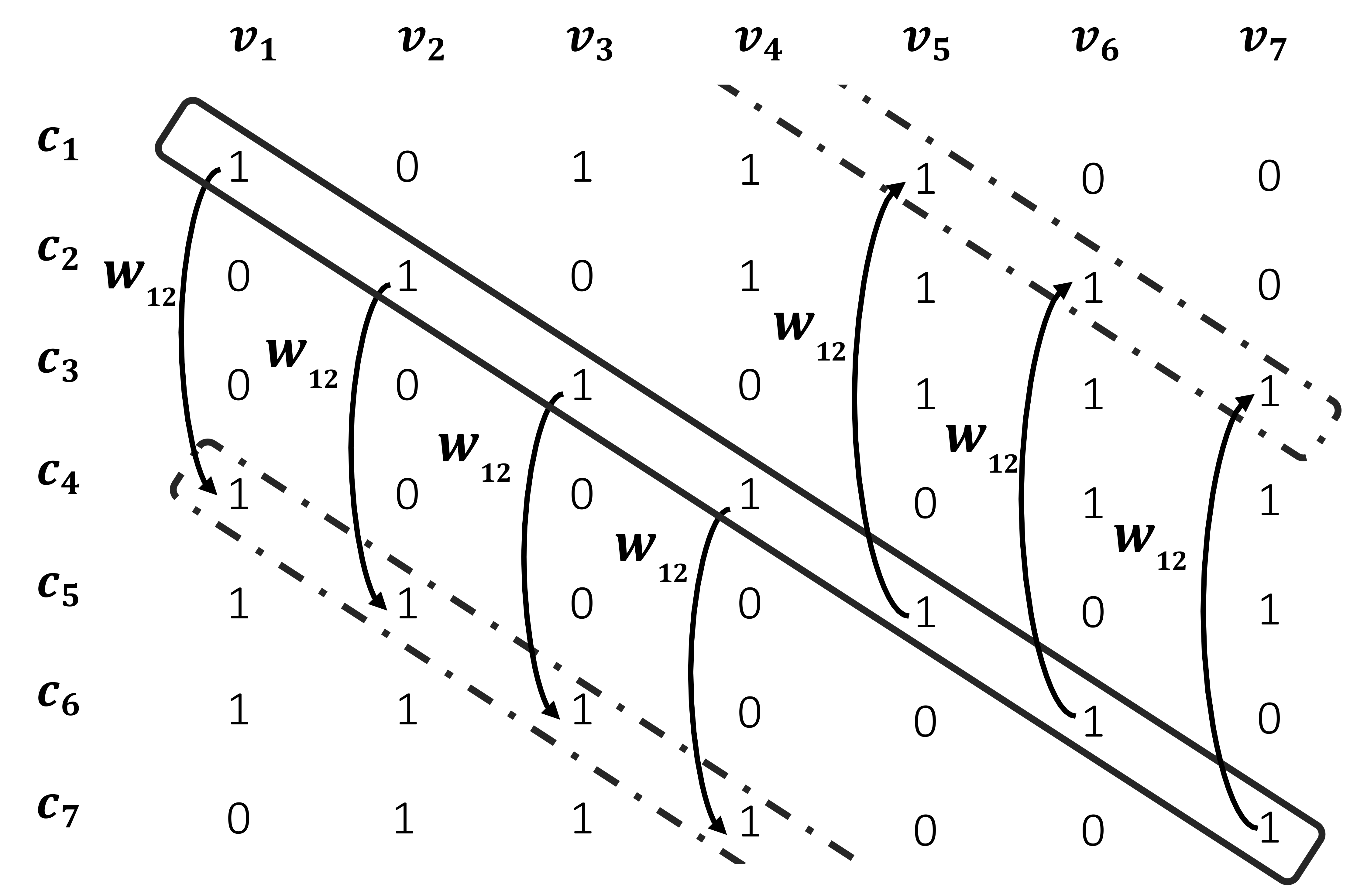} 
\end{subfigure}
\hspace*{1in}
\begin{subfigure}{0.16\textwidth}
\centering
\includegraphics[width=\textwidth]{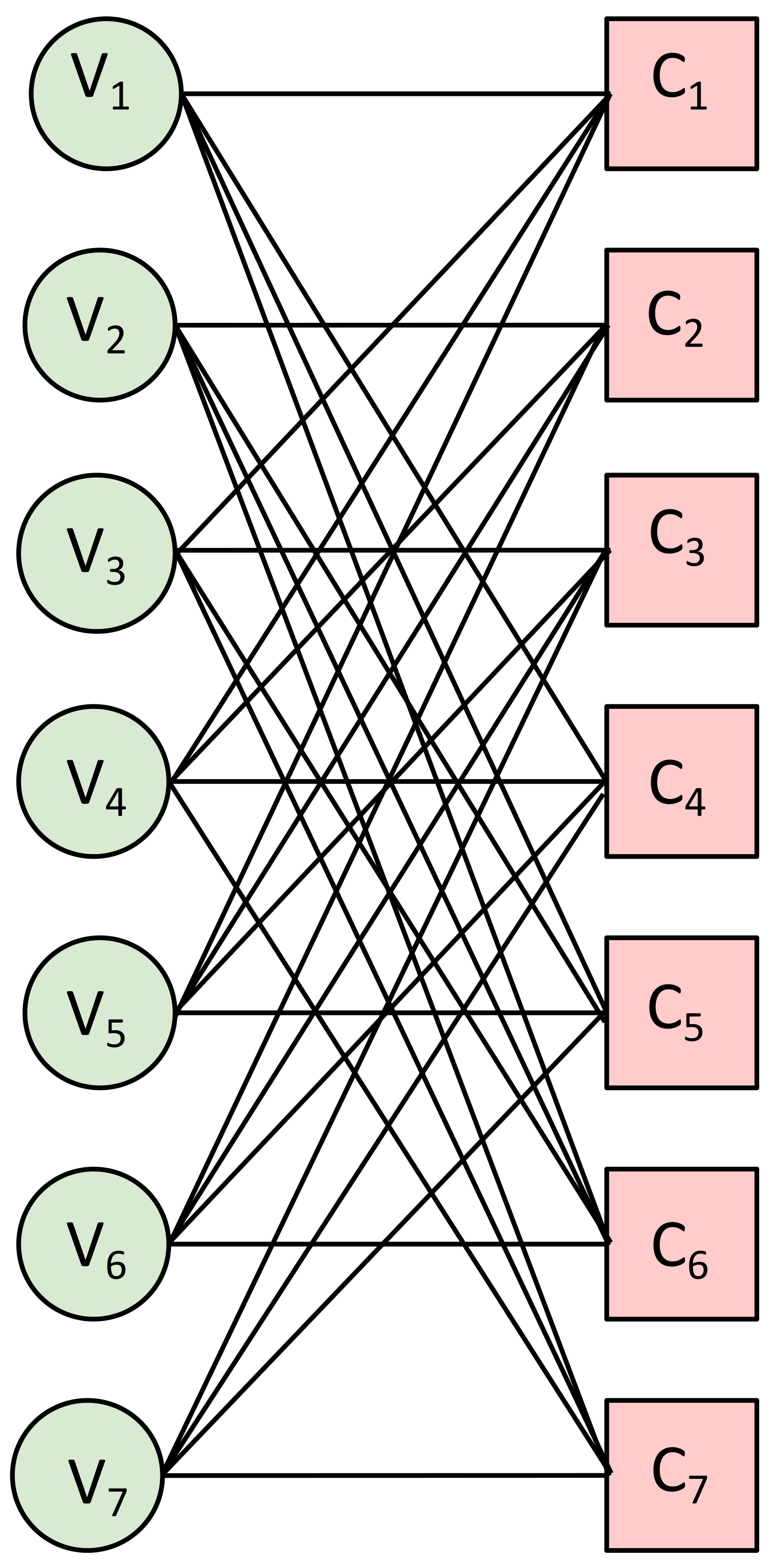} 
\end{subfigure}
\caption{The parity check matrix of the $(7,4)$ Hamming code used in our decoder (left) and the corresponding Tanner graph (right). Each column of the matrix is obtained by one cyclic downward shift of its previous column.
The weight $w^{[s]}((c_{\pi_j(1)},v_j), (c_{\pi_j(4)},v_j))$ in \eqref{eq:weightedodd} can be viewed as a measure of ``importance" of $x^{[s-1]}((c_{\pi_j(1)},v_j))$ in the calculation of $x^{[s]}((c_{\pi_j(4)},v_j))$, where $\pi_j$ is defined in \eqref{eq:permu}.
The cyclically invariant structure implies that the relation between the edges $(c_{\pi_j(1)},v_j)$ and $(c_{\pi_j(4)},v_j)$ is the same for all $j\in[n]$, so in our decoder we set $w^{[s]}((c_{\pi_j(1)},v_j), (c_{\pi_j(4)},v_j))=w_{1,2}^{(s)}$ for all $j\in[n]$. The superscript of $w_{1,2}^{(s)}$ is omitted in the figure.
}
\label{fig:our_matrix}
\end{figure*}

\begin{figure*}
\centering
\includegraphics[width=\textwidth]{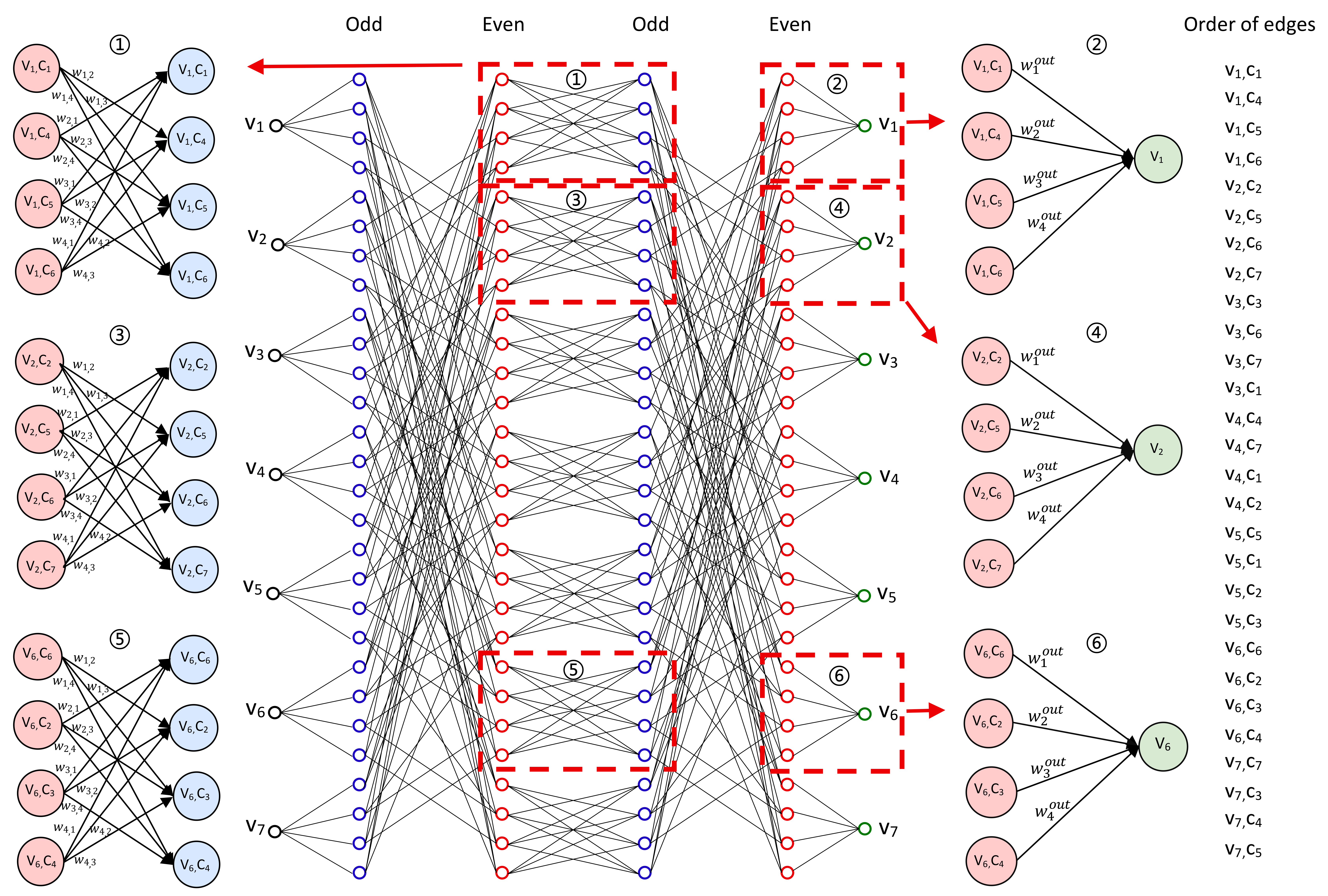}
\caption{The Trellis graph corresponding to the matrix in Fig.~\ref{fig:our_matrix}. The order of edges in the middle layers is listed in the rightmost column. Due to the cyclically invariant property, each $\big( x^{[s]}((c_{\pi_j(1)},v_j)), x^{[s]}((c_{\pi_j(4)},v_j)), x^{[s]}((c_{\pi_j(5)},v_j)), x^{[s]}((c_{\pi_j(6)},v_j)) \big)$ is the {\bf same} linear function of $\big( x^{[s-1]}((c_{\pi_j(1)},v_j)), x^{[s-1]}((c_{\pi_j(4)},v_j)), x^{[s-1]}((c_{\pi_j(5)},v_j)), x^{[s-1]}((c_{\pi_j(6)},v_j)) \big)$ and $L_j$ for all $j\in[7]$, meaning that the {\bf same} set of weights is shared among all $j\in[7]$ in \eqref{eq:weightedodd}. This is illustrated by the small figures \textcircled{\raisebox{-0.9pt}{1}},\textcircled{\raisebox{-0.9pt}{3}},\textcircled{\raisebox{-0.9pt}{5}}. (Weight of $L_j$ is omitted.) Similarly, in the output layer, the {\bf same} set of weights $w_1^{\out},w_2^{\out},w_3^{\out},w_4^{\out}$ is shared among all the $7$ linear mappings from $\big( x^{[2t]}((c_{\pi_j(1)},v_j)), x^{[2t]}((c_{\pi_j(4)},v_j)), x^{[2t]}((c_{\pi_j(5)},v_j)), x^{[2t]}((c_{\pi_j(6)},v_j)) \big)$ to $o_j$ for all $j\in[7]$, as illustrated by figures \textcircled{\raisebox{-0.9pt}{2}},\textcircled{\raisebox{-0.9pt}{4}},\textcircled{\raisebox{-0.9pt}{6}}.}
\label{fig:our_trellis}
\end{figure*}

\section{Cyclic codes and our new decoder}
\label{sect:our}

A cyclic code $\mathcal{C}$ is a linear code satisfying the following property: If $(C_1,C_2,\dots,C_n)\in\mathcal{C}$ is a codeword, then its cyclic shift $(C_n,C_1,C_2,\dots,C_{n-1})$ is also a codeword in $\mathcal{C}$. This implies that cyclic codes are invariant under all cyclic shifts. More precisely, for $b\in[n]$, let us define the cyclic shift $\pi_b$ on the set $[n]$ as
\begin{equation} \label{eq:permu}
\begin{aligned}
& \pi_b(i) = i+b-1 \quad \text{for~~} 1\le i\le n-b+1 , \\
& \pi_b(i) = i+b-1-n \quad \text{for~~} n-b+2 \le i\le n .
\end{aligned}
\end{equation}
Then for every cyclic code $\mathcal{C}$ and every $b\in[n]$, 
\begin{align}
& \{(C_{\pi_b(1)}, C_{\pi_b(2)}, \dots, C_{\pi_b(n)}) : (C_1,C_2,\dots,C_n)\in\mathcal{C} \} \nonumber \\
& = \mathcal{C} .  \label{eq:cyin}
\end{align}

Now consider the Maximum Likelihood (ML) decoder of a cyclic code $\mathcal{C}$. By definition, given the channel outputs $y=(y_1,\dots,y_n)$, the ML decoding result $\hat{C}^{\ML}(y) = (\hat{C}_1, \dots, \hat{C}_n)$ satisfies that
$$
\prod_{j=1}^n \mathbb{P}(y_j|\hat{C}_j)
\ge \prod_{j=1}^n \mathbb{P}(y_j|C_j)
\text{~for all~} (C_1,\dots,C_n)\in\mathcal{C}.
$$
Therefore, for every $b \in [n]$,
\begin{align*}
\prod_{j=1}^n \mathbb{P}(y_{\pi_b(j)}|\hat{C}_{\pi_b(j)})
\ge & \prod_{j=1}^n \mathbb{P}(y_{\pi_b(j)}|C_{\pi_b(j)}) \\
& \text{for all~} (C_1,\dots,C_n)\in\mathcal{C}.
\end{align*}
The cyclically invariant property \eqref{eq:cyin} further implies that
\begin{align*}
\prod_{j=1}^n \mathbb{P}(y_{\pi_b(j)}|\hat{C}_{\pi_b(j)})
\ge & \prod_{j=1}^n \mathbb{P}(y_{\pi_b(j)}|C_j) \\
& \text{for all~} (C_1,\dots,C_n)\in\mathcal{C}.
\end{align*}
Thus the ML decoding result of $(y_{\pi_b(1)},\dots,y_{\pi_b(n)})$ is
$$
\hat{C}^{\ML}\big( (y_{\pi_b(1)},\dots,y_{\pi_b(n)})
\big) = \big( \hat{C}_{\pi_b(1)}, \dots, \hat{C}_{\pi_b(n)} \big)
$$
for all $b \in [n]$. This proves that {\bf the ML decoder of a cyclic code is equivariant to all cyclic shifts}.

In light of this, we use a parity check matrix of size $n\times n$ instead of $(n-k)\times n$, and we impose a shift-invariant structure on the weights of our new decoder so that it shares the equivariant property of the ML decoder.

The first new ingredient of our decoder is the choice of the parity check matrix. A careful inspection of \eqref{eq:pc74} indicates that the rows of this parity check matrix are cyclic shifts of each other. This is in fact not a coincidence. According to Chapter~7.4 of \cite{Macwilliams77}, every cyclic code $\mathcal{C}$ with length $n$ and dimension $k$ possesses an $(n-k)\times n$ parity check matrix of the form
\begin{equation} \label{eq:generalpc}
\begin{array}{ccccccccc}
h_k & \dots & h_2 & h_1 & h_0 & 0 & 0 &\dots & 0 \\
0 & h_k & \dots & h_2 & h_1 & h_0 & 0 & \dots & 0 \\
\vdots & \vdots & \vdots & \vdots & \vdots & \vdots & \vdots & \vdots & \vdots \\
0 & 0 & \dots & 0 & h_k & \dots & h_2 & h_1 & h_0
\end{array} ,
\end{equation}
where $(h_k,\dots,h_1,h_0)$ is a binary vector of length $k+1$. For the $(n=7,k=4)$ Hamming code, the parity check matrix in \eqref{eq:pc74} corresponds to $(h_4,h_3,h_2,h_1,h_0)=(1,0,1,1,1)$. It is well known that the dual code of a cyclic code is also cyclic; see Chapter~7.4 of \cite{Macwilliams77}. As a consequence, all the cyclic shifts of the rows in \eqref{eq:generalpc} are parity checks of the code $\mathcal{C}$. In our new decoder, we use an $n\times n$ parity check matrix consisting of all the $n$ cyclic shifts of the first row of matrix \eqref{eq:generalpc}, as opposed to matrix \eqref{eq:generalpc} itself, which only contains the first $(n-k)$ cyclic shifts of its first row. Note that matrix \eqref{eq:generalpc} is widely used in previous neural decoders when decoding cyclic codes such as BCH codes \cite{Nachmani16,Nachmani18}.

As a concrete example, for the $(n=7,k=4)$ Hamming code, the matrix in Fig.~\ref{fig:our_matrix} is used in our new decoder. Each row in this matrix is obtained by one cyclic right shift of its previous row. As a consequence, each column is also obtained by one cyclic downward shift of its previous column.
According to \eqref{eq:weightedodd}, only the edges sharing the same variable node are involved in the calculations of odd layers. More specifically, for the Tanner graph in Fig.~\ref{fig:our_matrix} and for odd $s$, the calculations of $\big( x^{[s]}((c_1,v_1)), x^{[s]}((c_4,v_1)), x^{[s]}((c_5,v_1)), x^{[s]}((c_6,v_1)) \big)$ only involve $\big( x^{[s-1]}((c_1,v_1)), x^{[s-1]}((c_4,v_1)), \linebreak[4]  x^{[s-1]}((c_5,v_1)), 
x^{[s-1]}((c_6,v_1)) \big)$ and $L_1$; calculations of $\big( x^{[s]}((c_2,v_2)), x^{[s]}((c_5,v_2)), x^{[s]}((c_6,v_2)), x^{[s]}((c_7,v_2)) \big)$ only involve $\big( x^{[s-1]}((c_2,v_2)), x^{[s-1]}((c_5,v_2)), \linebreak[4] x^{[s-1]}((c_6,v_2)), x^{[s-1]}((c_7,v_2)) \big)$ and $L_2$; and so on. The weight $w^{[s]}((c_1,v_1), (c_4,v_1))$ in \eqref{eq:weightedodd} can be viewed as a measure of ``importance" of $x^{[s-1]}((c_1,v_1))$ in the calculation of $x^{[s]}((c_4,v_1))$. Similarly, $w^{[s]}((c_2,v_2), (c_5,v_2))$ measures the ``importance" of $x^{[s-1]}((c_2,v_2))$ in the calculation of $x^{[s]}((c_5,v_2))$. An important observation is that due to the cyclically invariant structure of the code and the parity matrix, the relation between the edges $(c_1,v_1)$ and $(c_4,v_1)$ is the same as the relation between $(c_2,v_2)$ and $(c_5,v_2)$. Therefore, in our decoder, we enforce the weights $w^{[s]}((c_1,v_1), (c_4,v_1))$ and $w^{[s]}((c_2,v_2), (c_5,v_2))$ to take the same value. More generally, we set
\begin{equation} \label{eq:share}
w^{[s]}((c_{\pi_j(1)},v_j), (c_{\pi_j(4)},v_j))
= w^{[s]}((c_1,v_1), (c_4,v_1))
\end{equation}
for all $j\in[n]$, as illustrated in Fig.~\ref{fig:our_matrix}, where the cyclic shift $\pi_j$ is defined in \eqref{eq:permu}.
In the calculations of $\big( x^{[s]}((c_1,v_1)), x^{[s]}((c_4,v_1)), x^{[s]}((c_5,v_1)), x^{[s]}((c_6,v_1)) \big)$, there are $4\times 3=12$ weights multiplied with $\big( x^{[s-1]}((c_1,v_1)), x^{[s-1]}((c_4,v_1)),   x^{[s-1]}((c_5,v_1)), \linebreak[4]
x^{[s-1]}((c_6,v_1)) \big)$. As a natural generalization of \eqref{eq:share}, in our decoder these 12 weights are shared in the calculations of $\big( x^{[s]}((c_{\pi_j(1)},v_j)), x^{[s]}((c_{\pi_j(4)},v_j)), x^{[s]}((c_{\pi_j(5)},v_j)), \linebreak[4] x^{[s]}((c_{\pi_j(6)},v_j)) \big)$ for every $j\in[n]$. In other words, we set
$$
w^{[s]}((c_{\pi_j(i_1)},v_j), (c_{\pi_j(i_2)},v_j))
= w^{[s]}((c_{i_1},v_1), (c_{i_2},v_1))
$$
for all $j\in[n]$ and all $i_1,i_2\in\{1,4,5,6\}$ such that $i_1\neq i_2$; see Fig.~\ref{fig:our_trellis} for an illustration.

In addition to the changes of $w^{[s]}(e',e)$ in \eqref{eq:weightedodd}, we also make changes to $w^{[s]}(j,e)$, the weight of $L_j$. Again by the cyclically invariant structure of the code and the matrix, the role of $L_1$ in the calculation of $x^{[s]}((c_1,v_1))$ is the same as the role of $L_2$ in the calculation of $x^{[s]}((c_2,v_2))$. Therefore, in our decoder, we enforce the weights $w^{[s]}(1,(c_1,v_1))$ and $w^{[s]}(2,(c_2,v_2))$ to take the same value. More generally, we set
$
w^{[s]}(j,(c_{\pi_j(i)},v_j))
=w^{[s]}(1,(c_i,v_1))
$
for all $j\in[n]$ and all $i\in\{1,4,5,6\}$.
Finally, in the calculations of the output layer, we also impose a similar shift-invariant structure on the weights $w^{\out}(e,j)$ in \eqref{eq:weightedout}. More precisely, we set
$
w^{\out}((c_{\pi_j(i)},v_j), j)
=w^{\out}((c_i,v_1), 1)
$
for all $j\in[n]$ and all $i\in\{1,4,5,6\}$.
see Fig.~\ref{fig:our_trellis} for an illustration.

Now we have finished the description of our new decoder for the special case of $(7,4)$ Hamming code. For a general cyclic code, the first step of our decoding algorithm is to extend the $(n-k) \times n$ parity check matrix in \eqref{eq:generalpc} to obtain an $n\times n$ parity check matrix $H$ of the form in Fig.~\ref{fig:our_matrix}. Suppose that the number of $1$'s in each column of $H$ is $u$, and let $\{i_1,i_2,\dots,i_u\}\subseteq [n]$ be the set satisfying $H_{i_1,1}=H_{i_2,1}=\dots=H_{i_u,1}=1$. Then $(c_{\pi_j(i_1)}, v_j), (c_{\pi_j(i_2)}, v_j), \dots, (c_{\pi_j(i_u)}, v_j)$ are the $u$ edges that contain $v_j$ as an endpoint in the Tanner graph. In the calculations of each odd layer, we use the following $u^2$ weights: $\{w_{b,b'}^{[s]}:b,b'\in[u],b\neq b'\}$ and $\{w_b^{[s]}:b\in[u]\}$. More precisely, for odd $s$ and an edge $e=(c_{\pi_j(i_b)}, v_j)$ with $b\in[u]$, the message $x^{[s]}(e)$ is given by
\begin{align}
 x^{[s]}(e)=
x^{[s]} & ((c_{\pi_j(i_b)}, v_j))  
= \tanh \Big( \frac{1}{2} \Big( w_b^{[s]} L_j
\label{eq:ourodd} \\
& + \sum_{b'\in [u]\setminus \{b\} }
w_{b', b}^{[s]} ~
x^{[s-1]}((c_{\pi_j(i_{b'})}, v_j)) \Big) \Big) .
\nonumber
\end{align}
In the calculations of the output layer, we use the following $u$ weights: $\{w_b^{\out}:b\in[u]\}$. The $j$th output is given by
\begin{equation} \label{eq:ourout}
o_j = L_j  +  \sum_{b\in[u]} 
w_b^{\out} ~
x^{[2t]}((c_{\pi_j(i_b)}, v_j)) 
\end{equation}
for $j\in[n]$, where $t$ is the total number of iterations. As for the calculations of even layers, we still use the same formula \eqref{eq:classiceven} as in the vanilla BP algorithm. In the supplementary material, we prove that {\bf our new decoder is equivariant to all cyclic shifts}.

\begin{table*}
    \centering
    \caption{The negative natural logarithm of BER for three SNR values and different decoders. Higher is better. N18 refers to \cite{Nachmani18}, and $B$ is the number of boosting used in the decoder. For all codes, we train with a batch size of $160$ samples, among which we produce $20$ samples from each of the following $8$ SNR values: 1dB, 2dB, \dots, 8dB. We use $10^5$ samples for testing.}
    \label{tab:ber}
    \vskip 0.15in
    \begin{center}
    \begin{small}
    
\begin{tabular}{l c@{~~}c@{~~}c c@{~~}c@{~~}c c@{~~}c@{~~}c c@{~~}c@{~~}c c@{~~}c@{~~}c}

    \toprule
        Decoder  & 
        \multicolumn{3}{c}{BP} & 
        \multicolumn{3}{c}{N18, $B=0$} & 
        \multicolumn{3}{c}{N18, $B=2$} & 
        \multicolumn{3}{c}{Ours, $B=0$} & 
        \multicolumn{3}{c}{Ours, $B=2$}\\

        \cmidrule(lr){2-4}
        \cmidrule(lr){5-7}
        \cmidrule(lr){8-10}
        \cmidrule(lr){11-13}
        \cmidrule(lr){14-16}
       Code/SNR  & 4 & 5 & 6 & 4 & 5 & 6 & 4 & 5 & 6 & 4 & 5 & 6 & 4 & 5 & 6\\ 
        \midrule 
        BCH(63,24)  & 3.18 & 4.07 & 5.18 & 3.24 & 4.36 & 5.80 & 3.43 & 4.62 & 6.16 & 3.78 & 5.11 & 7.07  & 3.98 & 5.42 & 7.34  \\
        BCH(63,36)  & 3.83 & 4.69 & 5.91 & 3.97 & 5.27 & 7.05 & 4.00 & 5.35 & 7.43 & 4.63 & 6.48 & 8.86  & 4.75 & 6.40 & 10.02 \\
        BCH(63,45)  & 3.94 & 4.84 & 6.30 & 4.37 & 5.71 & 7.45 & 4.42 & 5.83 & 7.50 & 5.12 & 6.97 & 9.46 & 5.39 & 7.45 & 10.45  \\
        BCH(127,36) & 2.21 & 2.71 & 3.52 & 2.23 & 2.76 & 3.72 & 2.31 & 3.06 & 4.38 & 2.29 & 2.96 & 4.24  & 2.44 & 3.36 & 4.98  \\
        BCH(127,64) & 2.99 & 3.63 & 4.33 & 3.00 & 3.82 & 5.01 & 3.04 & 3.91 & 5.29 & 3.17 & 4.24 & 6.05  & 3.23 & 4.43 & 6.42  \\
        BCH(127,99) & 3.67 & 4.38 & 5.87 & 4.08 & 5.34 & 7.12 & 4.07 & 5.42 & 7.29 & 4.51 & 6.30 & 8.86 & 4.63 & 6.59 & 9.61 \\
        PRM(63,22)   & 2.83 & 3.55 & 4.46 & 2.85 & 3.76 & 4.98 & 3.07 & 3.99 & 5.33 & 3.32 & 4.52 & 6.23  & 3.70 & 5.13 & 7.16  \\
        PRM(63,42)   & 4.61 & 6.00 & 7.79 & 4.81 & 6.47 & 8.87 & 4.85 & 6.47 & 9.11 & 5.92 & 8.26 & 10.85 & 6.27 & 8.81 & 11.55 \\
        PRM(127,64)  & 2.95 & 3.53 & 4.12 & 2.89 & 3.55 & 4.90 & 2.94 & 3.69 & 5.00 & 3.14 & 4.17 & 5.93  & 3.22 & 4.37 & 6.35  \\
        PRM(127,99)  & 4.44 & 5.82 & 7.65 & 4.56 & 6.30 & 8.35 & 4.58 & 6.33 & 8.43 & 5.69 & 8.19 & 11.15 & 5.97 & 8.83 & 14.33 \\
        \bottomrule
    \end{tabular}
\end{small}
\end{center}
\vskip -0.1in
\end{table*}

\section{List decoding procedure}
\label{sect:list}
BCH codes and punctured RM codes are two families of extensively studied and widely applied cyclic codes. In this section we propose a list decoding procedure for these two code families. Let $(C_1,C_2,\dots,C_n)$ be a codeword of an $(n,k)$ BCH code. If we prepend an overall parity bit $C_0=C_1+C_2+\dots+C_n$ to this codeword (addition is over binary field), then the resulting vector $(C_0,C_1,C_2,\dots,C_n)$ is a codeword of an $(n+1,k)$ extended BCH code. Similarly, prepending an overall parity bit to an $(n,k)$ punctured RM code results in an $(n+1,k)$ RM code. Both RM codes and extended BCH codes are invariant to the affine group \cite{Kasami67}. In the list decoding procedure, we use the permutations from a subset $\mathcal{S}$ of the affine group. The details about how to choose this subset $\mathcal{S}$ are elaborated in the supplementary material. For now we only need to know that the size of $\mathcal{S}$ is equal to the length of the corresponding RM or extended BCH code.

We will describe our list decoding algorithm for BCH codes, and the algorithm for punctured RM codes is exactly the same. Let $(L_1,L_2,\dots,L_n)$ be the LLR vector of the channel outputs after transmitting a randomly chosen codeword $(C_1,\dots,C_n)$ from a BCH code with code length $n$. Then the size of $\mathcal{S}$ is $|\mathcal{S}|=n+1$. Choose the list size in the list decoding algorithm to be $\ell\le n+1$, and the algorithm works as follows:\\
{\bf Step 1: Prepend a dummy symbol $L_0=0$ to the LLR vector.} As mentioned above, $(C_0,C_1,\dots,C_n)$ is a codeword from an extended BCH code, where $C_0$ is the overall parity. The dummy symbol $L_0=0$ is the LLR of $C_0$, indicating that the probability of $C_0=0$ is the same as the probability of $C_0=1$ because we do not have any direct information about $C_0$.\\
{\bf Step 2: Apply $\ell$ permutations to $(L_0,L_1,\dots,L_n)$.} We pick $\ell$ permutations from the set $\mathcal{S}$. Denote them as $\sigma_1,\dots,\sigma_\ell$. Apply these permutations to $(L_0,L_1,\dots,L_n)$ and we obtain $\ell$ vectors $(L_0^{(i)},L_1^{(i)},\dots,L_n^{(i)})=(L_{\sigma_i(0)},L_{\sigma_i(1)},\dots,L_{\sigma_i(n)})$ for $i\in[\ell]$.\\
{\bf Step 3: Decode $(L_1^{(i)},L_2^{(i)},\dots,L_n^{(i)})$ using our neural decoder.} For each $i\in[\ell]$, decode $(L_1^{(i)},L_2^{(i)},\dots,L_n^{(i)})$ using our decoder proposed in Section~\ref{sect:our} and denote the result as $(\bar{C}_1^{(i)},\bar{C}_2^{(i)},\dots,\bar{C}_n^{(i)})$.\\
{\bf Step 4: Check whether $(\bar{C}_1^{(i)},\bar{C}_2^{(i)},\dots,\bar{C}_n^{(i)})$ is a codeword or not.} Multiply $(\bar{C}_1^{(i)},\bar{C}_2^{(i)},\dots,\bar{C}_n^{(i)})$ with the parity check matrix of the BCH code. If the result is $0$ (meaning that this is a codeword) then we do nothing; otherwise we set $(\bar{C}_1^{(i)},\bar{C}_2^{(i)},\dots,\bar{C}_n^{(i)})$ to be the all zero vector.\\
{\bf Step 5: Prepend an overall parity to $(\bar{C}_1^{(i)},\dots,\bar{C}_n^{(i)})$.} Prepend an overall parity $\bar{C}_0^{(i)}=\bar{C}_1^{(i)}+\bar{C}_2^{(i)}+\dots+\bar{C}_n^{(i)}$ to obtain $(\bar{C}_0^{(i)},\bar{C}_1^{(i)},\dots,\bar{C}_n^{(i)})$.\\
{\bf Step 6: Apply the inverse permutation $\sigma_i^{-1}$ to $(\bar{C}_0^{(i)},\bar{C}_1^{(i)},\dots,\bar{C}_n^{(i)})$.} Apply the inverse permutation $\sigma_i^{-1}$ to obtain $(\hat{C}_0^{(i)},\hat{C}_1^{(i)},\dots,\hat{C}_n^{(i)})= (\bar{C}_{\sigma_i^{-1}(0)}^{(i)},\bar{C}_{\sigma_i^{-1}(1)}^{(i)},\dots,\bar{C}_{\sigma_i^{-1}(n)}^{(i)})$ for all $i\in[\ell]$.\\
{\bf Step 7: ML decoding among the $\ell$ candidates $(\hat{C}_0^{(i)},\hat{C}_1^{(i)},\dots,\hat{C}_n^{(i)}), i\in[\ell]$.} ML decoding amounts to finding $i\in[\ell]$ that minimizes $L_0 \hat{C}_0^{(i)}+L_1 \hat{C}_1^{(i)}+\dots+L_n \hat{C}_n^{(i)}$.  Denote the minimizer as $i^*$ and set $(\hat{C}_0,\hat{C}_1,\dots,\hat{C}_n)=(\hat{C}_0^{(i^*)},\hat{C}_1^{(i^*)},\dots,\hat{C}_n^{(i^*)})$.\\
{\bf Step 8: Discard the first bit $\hat{C}_0$}. The final decoding result is $(\hat{C}_1,\hat{C}_2,\dots,\hat{C}_n)$.

Note that in Step 3, our neural decoder can be replaced by any decoder of BCH codes. As a final remark, Step 4 is very important for reducing the decoding error probability because without this step, some non-codeword might be the minimizer in Step 7, which results in a decoding error.

\begin{table*}
    \centering
\caption{The negative natural logarithm of FER of the list decoding algorithm. $>11.5$ means FER$<10^{-5}$ since we only use $10^5$ samples for testing. For all list sizes, our decoder is boosted $20$ times for codes with length $63$ and $50$ times for codes with length $127$.}
\label{tab:list}
\vskip 0.15in
\begin{center}
\begin{small}
    \begin{tabular}{lc@{~~}c@{~~}c c@{~~}c@{~~}c c@{~~}c@{~~}c c@{~~}c@{~~}c c@{~~}c@{~~}c}
    
    \toprule
        
       List size & 
        \multicolumn{3}{c}{$\ell=1$} &
        \multicolumn{3}{c}{$\ell=2$} &  
        \multicolumn{3}{c}{$\ell=4$} & 
        \multicolumn{3}{c}{$\ell=8$} & \multicolumn{3}{c}{$\ell=n+1$}\\
        
        \cmidrule(lr){2-4}
        \cmidrule(lr){5-7}
        \cmidrule(lr){8-10}
        \cmidrule(lr){11-13}
        \cmidrule(lr){14-16}
    Code/SNR    & 4 & 5 & 6 & 4 & 5 & 6 & 4 & 5 & 6 & 4 & 5 & 6 & 4 & 5 & 6\\ 
        \midrule 
    BCH(63,24)  & 2.51 & 4.02 & 6.21 & 3.30 & 5.08 & 7.60 & 4.40 & 6.21 & $>$11.5 & 5.52 & 9.21 & $>$11.5 & 9.21 & $>$11.5 & $>$11.5 \\
    BCH(63,36)  & 2.89 & 4.68 & 7.49      & 3.53 & 5.65 & 8.87                & 4.23 & 7.26 & 10.13               & 5.04 & 8.52 & $>$11.5 & 9.21 & $>$11.5 & $>$11.5 \\
    BCH(63,45)  & 3.01 & 5.07 & 8.15 & 3.48 & 5.76 & 8.42 & 4.01 & 6.68 & 9.90 & 4.65 & 7.49 & 10.41 & 6.06 & 9.57 & $>$11.5 \\
    BCH(127,36) & 0.62 & 1.50 & 3.13  & 0.94 & 2.12 & 4.26 & 1.35 & 2.92 & 5.71 & 1.92 & 3.95 & 7.26 & 5.10 & 9.21 & $>$11.5 \\
    BCH(127,64)   & 0.79 & 2.01 & 4.00     & 1.10 & 2.61 & 5.08                & 1.51 & 3.41 & 6.34                & 2.02 & 4.17 & 7.52                & 4.66 & 7.82                & $>$11.5 \\
    BCH(127,99)  & 1.63 & 3.55 & 7.13  & 1.91 & 4.12 & 7.82 & 2.22 & 4.77 & $>$11.5  & 2.60 & 5.45 & $>$11.5  & 4.29 & 9.21 & $>$11.5 \\
    PRM(63,22)  & 2.03 & 3.45 & 5.43    & 2.70 & 4.40 & 7.13                & 3.61 & 5.68 & 8.11                & 4.79 & 7.82 & $>$11.5 & 9.21 & $>$11.5  & $>$11.5 \\
    PRM(63,42)  & 4.06 & 6.57 & 9.90     & 4.55 & 7.39 & $>$11.5 & 5.18 & 8.62 & $>$11.5 & 5.91 & 9.21 & $>$11.5 & 6.21 & 9.72                & $>$11.5 \\
    PRM(127,64)  & 0.69 & 1.83 & 3.82      & 0.95 & 2.38 & 4.84                & 1.33 & 3.17 & 6.07                & 1.75 & 3.95 & 7.13                & 4.34 & 7.13                & $>$11.5 \\
    PRM(127,99)  & 3.13 & 6.15 & $>$11.5      & 3.61 & 6.95 & $>$11.5 & 4.06 & 7.82 & $>$11.5 & 4.41 & 8.42 & $>$11.5 & 5.50 & $>$11.5 & $>$11.5 \\
        \bottomrule
    \end{tabular}
\end{small}
\end{center}
\vskip -0.1in
\end{table*}

\begin{table*}
    \centering
    \caption{Ablation analysis. The numbers are negative natural logarithm of BER. N18 refers to \cite{Nachmani18}.}
    \label{tab:ablation}
    \vskip 0.15in
    \begin{center}
    \begin{small}
    \begin{tabular}{l c@{~~}c@{~~}c c@{~~}c@{~~}c c@{~~}c@{~~}c c@{~~}c@{~~}c}

    \toprule
        Code  & 
        \multicolumn{3}{c}{BCH(63,24)} & 
        \multicolumn{3}{c}{BCH(63,36)} & 
        \multicolumn{3}{c}{BCH(63,45)} & 
        \multicolumn{3}{c}{PRM(63,22)}\\

        \cmidrule(lr){2-4}
        \cmidrule(lr){5-7}
        \cmidrule(lr){8-10}
        \cmidrule(lr){11-13}

 Decoder/SNR       & 4 & 5 & 6 & 4 & 5 & 6 & 4 & 5 & 6 & 4 & 5 & 6\\
        \midrule 
        
        N18, $(n-k)\times n$ parity matrix & 3.24 & 4.36 & 5.80 & 3.97 & 5.27 & 7.05 & 4.37 & 5.71 & 7.45 & 2.85 & 3.76 & 4.98 \\
        N18, randomly extended $n\times n$ matrix        & 3.27 & 4.38 & 5.78 & 3.97 & 5.28 & 7.07 & 4.49 & 5.81 & 7.47 & 2.90 & 3.86 & 5.17 \\
        N18, $n\times n$ cyclic matrix    & 3.67 & 5.08 & 6.78 & 4.42 & 5.93 & 7.84 & 4.87 & 6.19 & 7.76 & 3.15 & 4.33 & 5.86 \\
        Ours, $n\times n$ cyclic matrix    & 3.78 & 5.11 & 7.07 & 4.63 & 6.48 & 8.86 & 5.12 & 6.97 & 9.46 & 3.32 & 4.52 & 6.23 \\
        \bottomrule
    \end{tabular}
\end{small}
\end{center}
\vskip -0.1in
\end{table*}

\section{Simulation results}
Similarly to the neural decoders proposed in \cite{Nachmani16,Nachmani18,Nachmani19}, it is easy to verify that our new decoder also satisfies the message passing symmetry conditions in Definition 4.81 of \cite{Richardson08}. Hence, by Lemma 4.90 of \cite{Richardson08}, the decoding error probability is independent of the transmitted codeword. A direct implication is that we can train our decoder solely using the all-zero codeword, instead of using randomly chosen codewords.

In the experiments, we find that a simple {\bf boosting method} can effectively reduce the decoding error probability. More precisely, the output vector of neural BP decoders (both ours and \cite{Nachmani16,Nachmani18}) is still an LLR vector (see \eqref{eq:weightedout} and \eqref{eq:ourout}), so we can feed this output LLR vector as an input to the neural decoder again, and we can repeat this procedure many times. 

\begin{figure*}
\centering
\begin{subfigure}{0.49\textwidth}
\centering
\includegraphics[width=\textwidth]{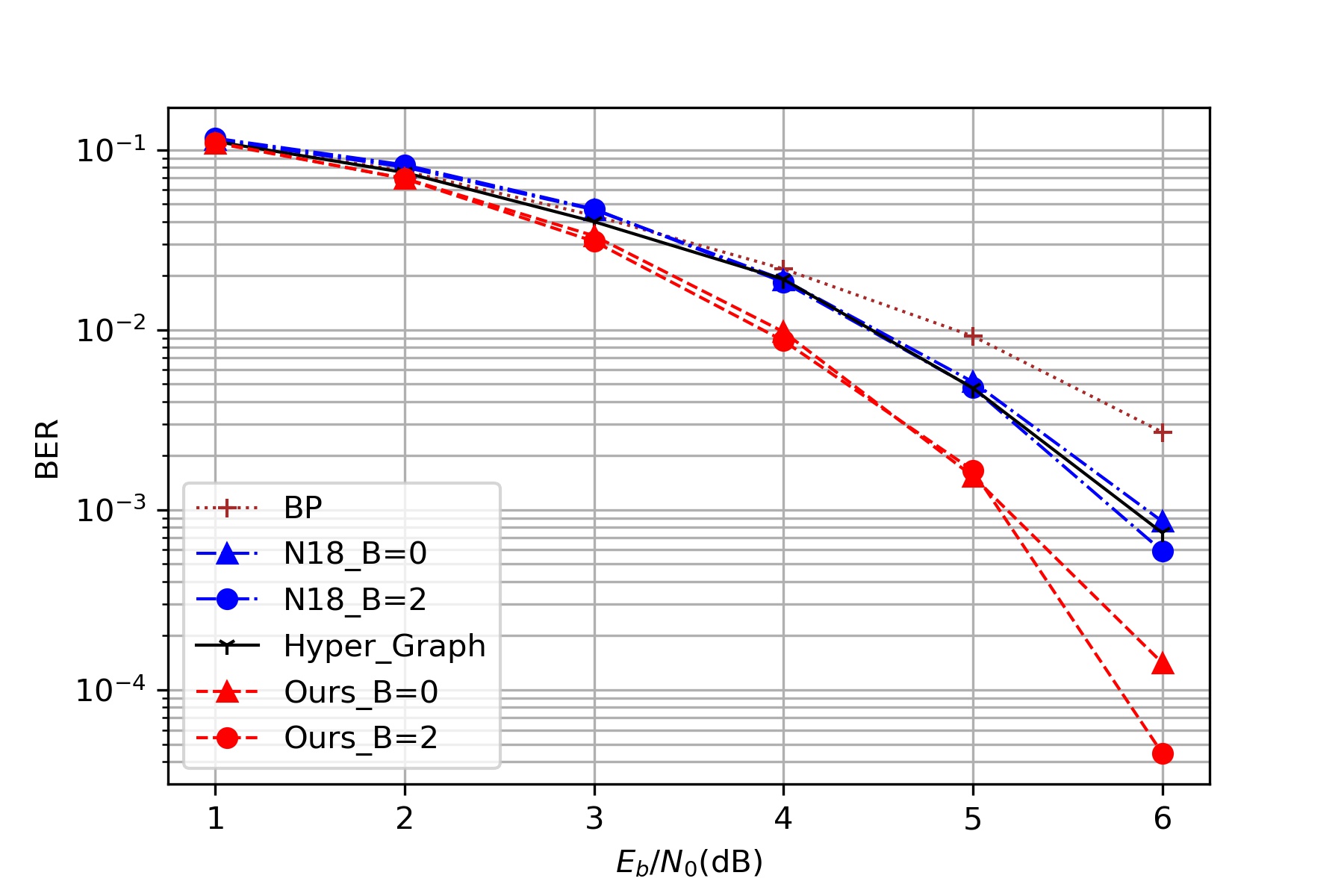} 
\caption{BCH(63,36)}
\label{fig:BCH63_36}
\end{subfigure}
~
\begin{subfigure}{0.49\textwidth}
\centering
\includegraphics[width=\textwidth]{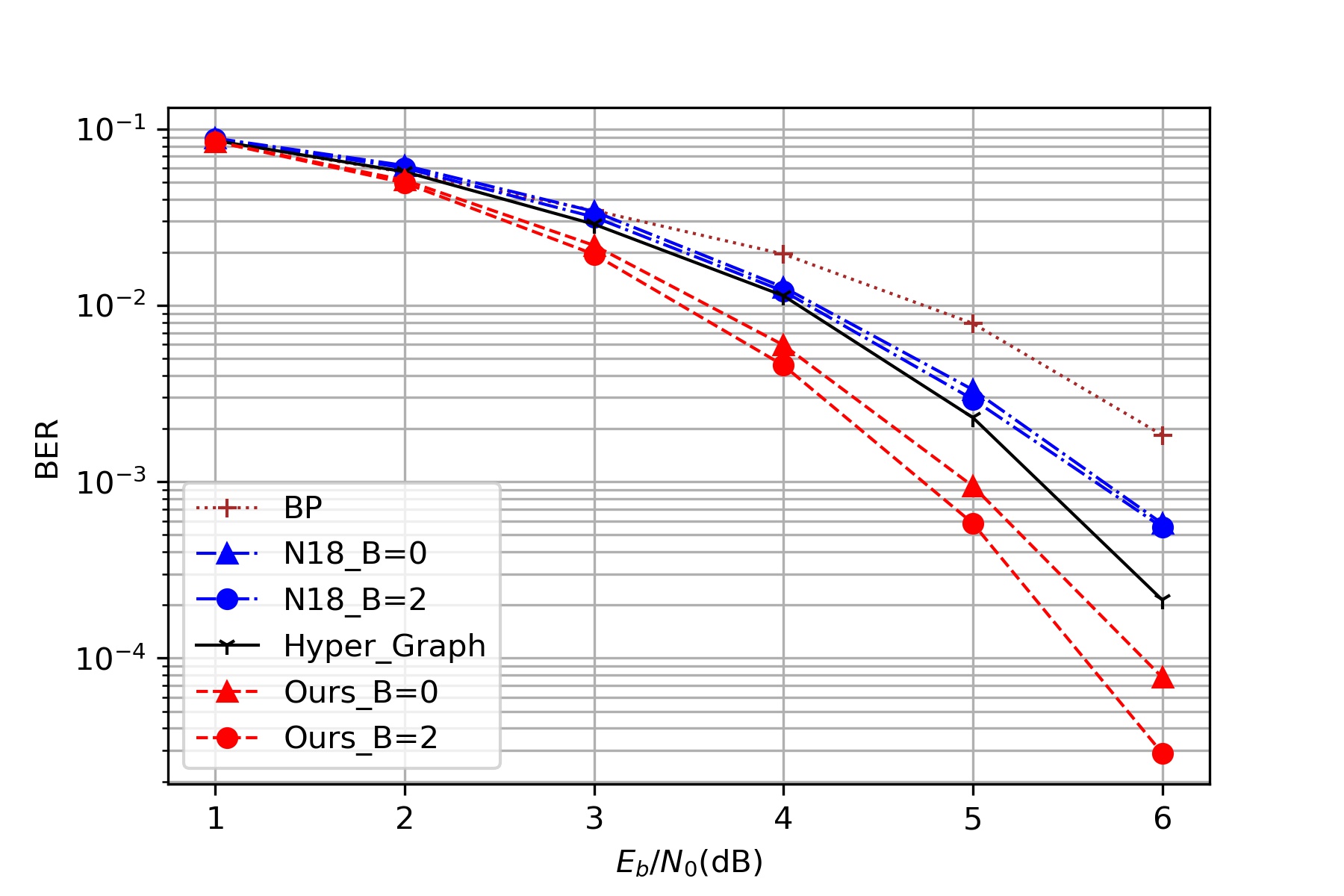} 
\caption{BCH(63,45)}
\label{fig:BCH63_45}
\end{subfigure}

\begin{subfigure}{0.49\textwidth}
\centering
\includegraphics[width=\textwidth]{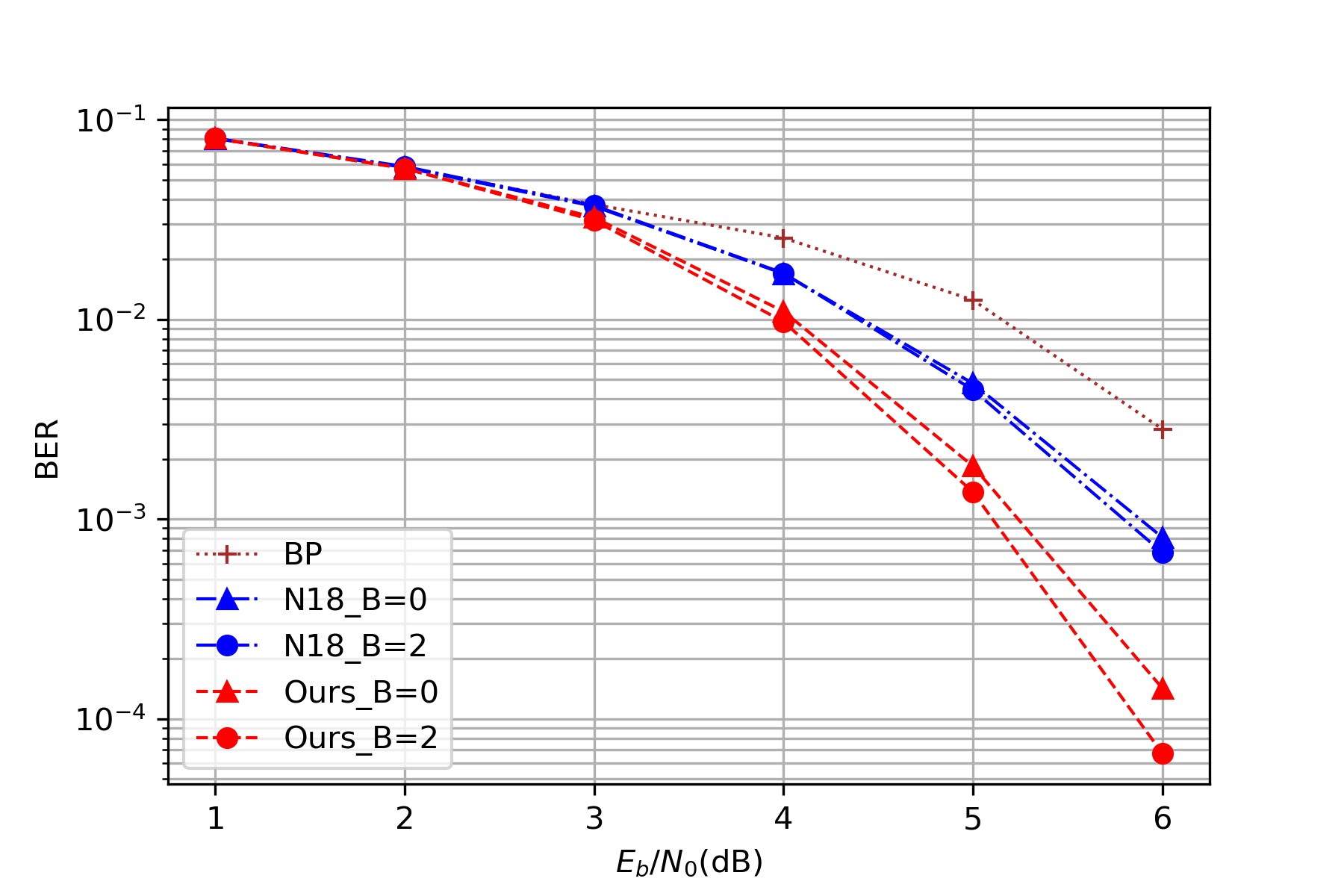} 
\caption{BCH(127,99)}
\end{subfigure}
~
\begin{subfigure}{0.49\textwidth}
\centering
\includegraphics[width=\textwidth]{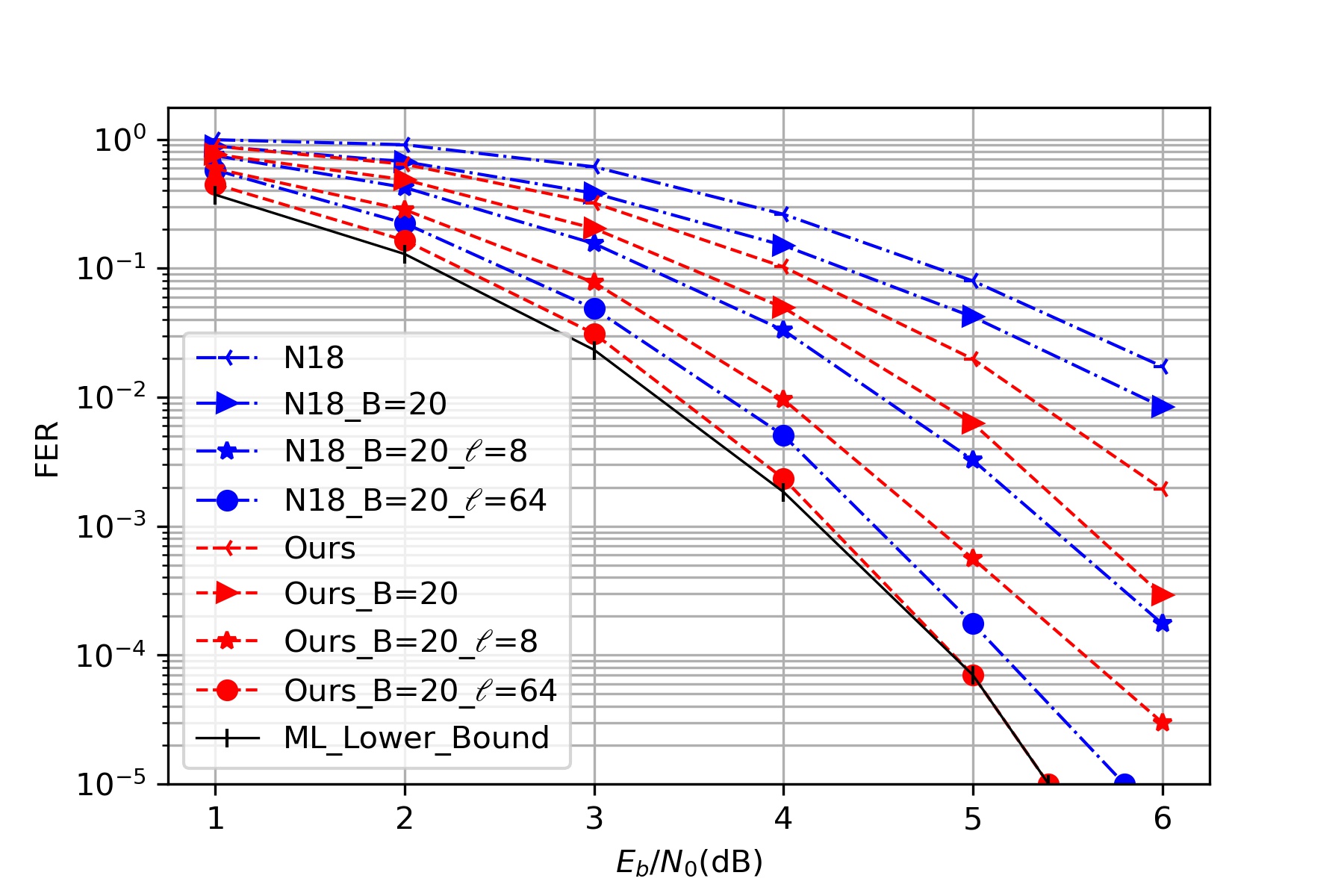} 
\caption{BCH(63,45) List decoding}
\end{subfigure}

\begin{subfigure}{0.49\textwidth}
\centering
\includegraphics[width=\textwidth]{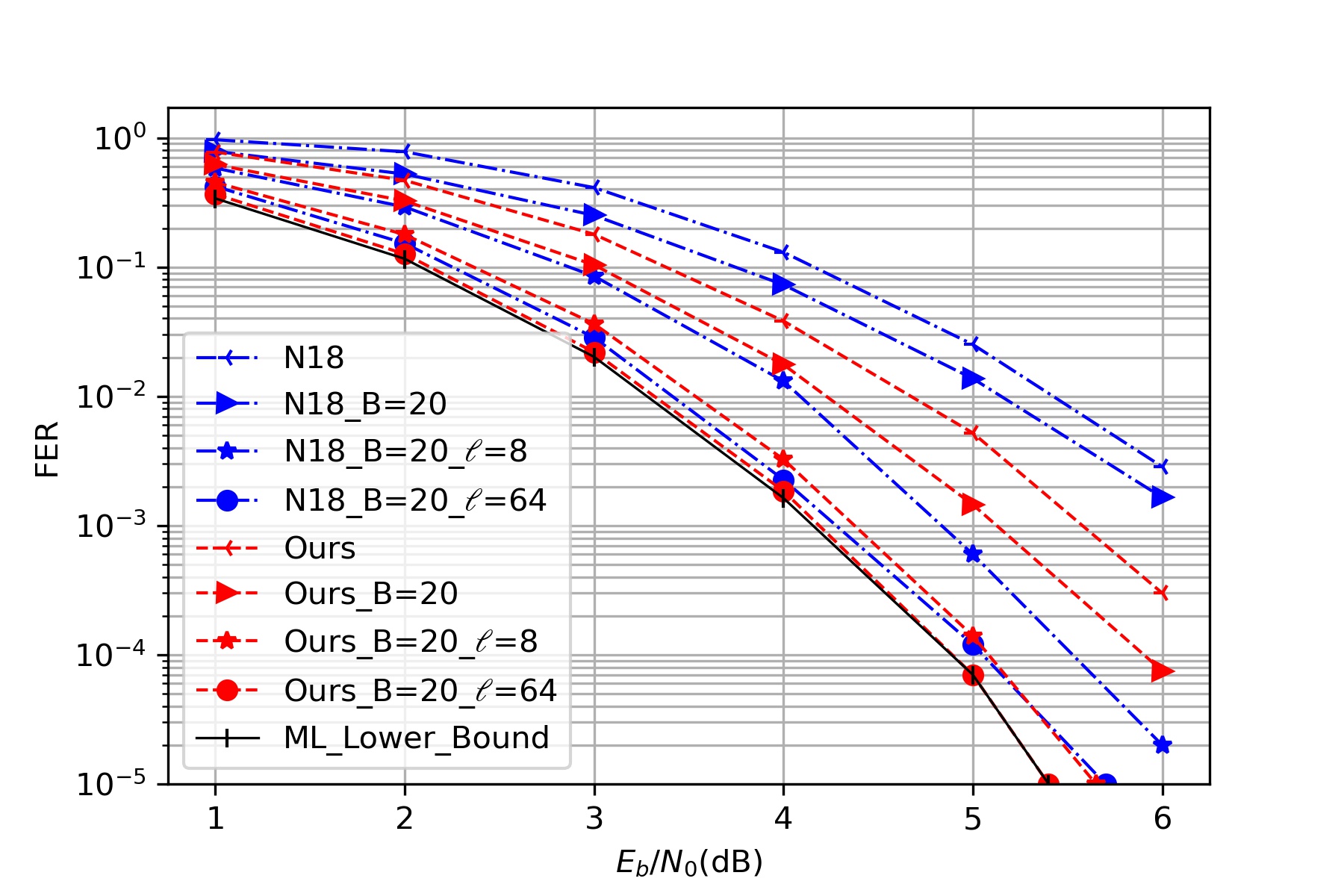} 
\caption{Punctured RM(63,42) List decoding}
\end{subfigure}
~
\begin{subfigure}{0.49\textwidth}
\centering
\includegraphics[width=\textwidth]{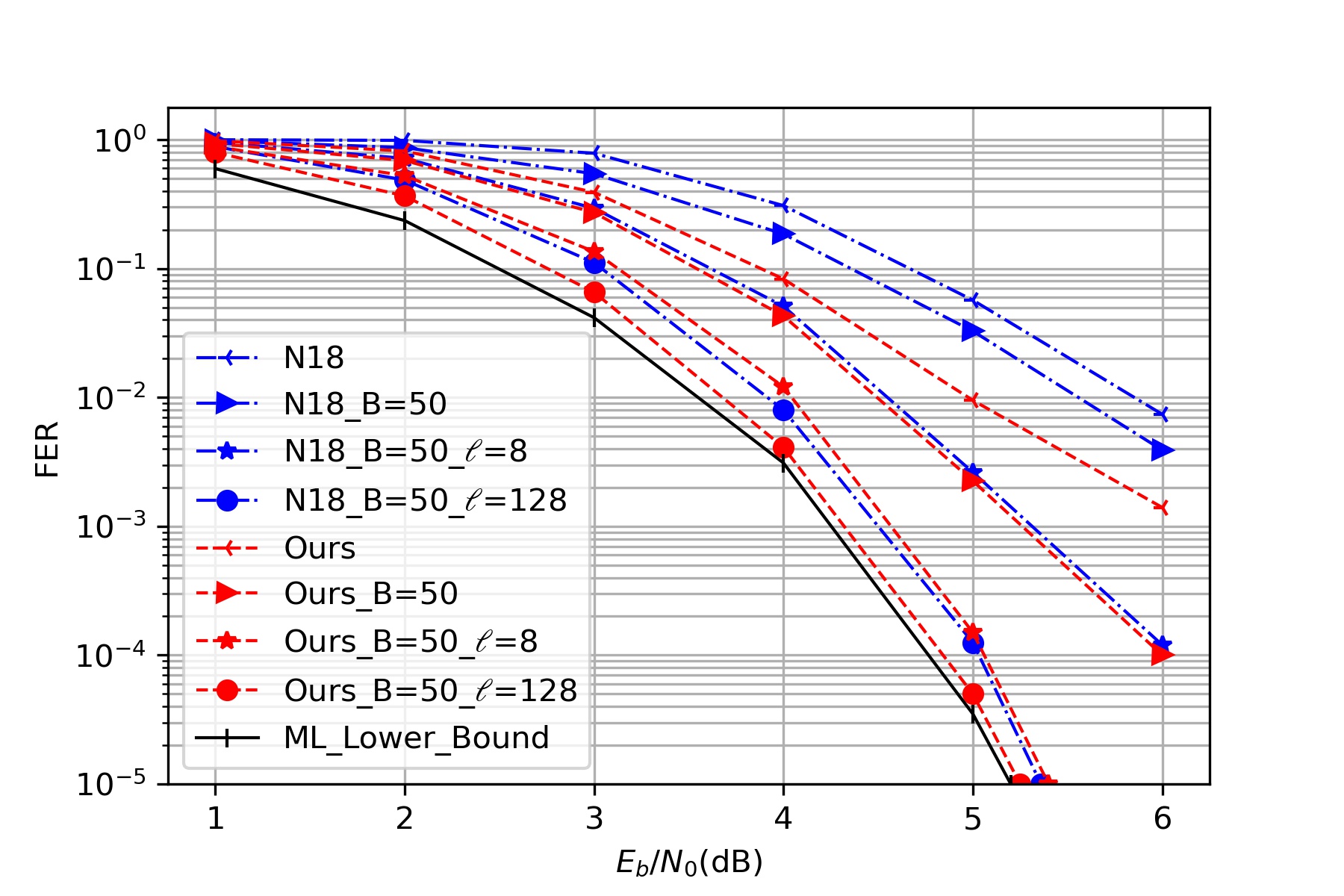} 
\caption{Punctured RM(127,99) List decoding}
\end{subfigure}

\caption{N18 refers to the neural decoder proposed in Section~III of \cite{Nachmani18}; $B$ is the number of boosting used in the decoder; $\ell$ is the list size. If we do not specify $B$ (respectively, $\ell$), it means that $B=0$ (respectively, $\ell=1$). For the first three plots, we use BER (the fraction of incorrect bits in the decoding results) to measure the decoding error probability, and for the last three plots, we use FER (the fraction of incorrect codewords in the decoding results) because it involves list decoding. Some additional plots are provided in the supplementary material.\\
Without list decoding, our neural decoder consistently improves upon \cite{Nachmani18} by $0.7$dB, and it also improves upon the hyper-graph-network decoder \cite{Nachmani19} by $0.3$dB. Moreover, the list decoding algorithm provides up to 3dB gain if we set the list size to be $n+1$. Even for a small list size $\ell=8$, it also gives $0.7\sim 0.9$dB gain over the algorithm without list decoding. In (d)--(f), when the list size is $\ell=n+1$, our list decoding algorithm has almost the same performance as the ML decoder.}
\label{fig:simu}
\end{figure*}

We compare the performance of our neural decoder with the feed-forward (FF) neural decoder proposed in Section~III of \cite{Nachmani18} and the hyper graph decoder in \cite{Nachmani19}. In \cite{Nachmani18}, several variations of the FF neural decoder were also proposed, such as RNN and min-sum decoders. Yet in many cases the FF neural decoder gives better performance than its variations in terms of decoding error probability, so we only compare with the FF decoder. For the hyper graph decoder, we are only able to obtain results for BCH$(63,36)$ and BCH$(63,45)$ due to the high complexity of its training procedure; see Fig.~\ref{fig:BCH63_36}--\ref{fig:BCH63_45}. We also test the performance of the list decoding algorithm with various list sizes. Following the common practice in the literature, we use Bit Error Rate (BER) to measure the decoding error probability for neural BP algorithms, and we use Frame Error Rate (FER) to measure the decoding error probability for list decoding algorithms. In our simulations, the number of BP iterations is $5$ for all the methods.

The simulation results are given in Tables~\ref{tab:ber}--\ref{tab:list} and Fig.~\ref{fig:simu}, and explanations of these results are given in the captions. Here we only discuss the comparison between our list decoder and the ML decoder: For certain high-rate codes (rate $\ge 0.7$), our list decoder with list size $\ell=n+1$ provides almost the same performance as the ML decoder. More precisely, simulation results in Fig.~\ref{fig:simu}(d)--(f) show that the gap between our list decoder with $\ell=n+1$ and the ML decoder is no larger than 0.1dB for these high-rate codes.
However, for codes with rate $\le 0.5$, the gap between our decoder with list size $\ell=n+1$ and the ML lower bound can be larger than 1dB. It remains unclear to us why there is such a distinction between high-rate codes and medium-to-low-rate codes.
As a final remark, we note that efficient decoders with almost ML performance were already proposed for low-rate RM codes; see for example \cite{Dumer06,Ye20}.

\begin{table}
   \centering
    \caption{Decoding time of N18 \cite{Nachmani18} and our decoder}
    \label{tb:time}
    \vskip 0.15in
        \begin{tabular}{cccc}
        \toprule
            Code & N18	& Ours	& N18/Ours \\
		\midrule 
            BCH(63,36)  & 4.09ms & 4.68ms & 0.87 \\
            BCH(63,45)  & 3.85ms & 4.94ms & 0.78 \\
            BCH(127,64) & 8.60ms & 8.63ms & 0.99 \\
            BCH(127,99) & 4.05ms & 16.7ms & 0.24 \\
            PRM(127,99)  & 3.97ms & 7.45ms & 0.53 \\
		\bottomrule 
        \end{tabular}
\end{table}

Next we compare the complexity between our decoder and \cite{Nachmani18}. The memory requirements of our decoder is much smaller than \cite{Nachmani18} while the time complexity of our decoder is larger than \cite{Nachmani18}.  In both our decoder and \cite{Nachmani18}, the memory is used to store the trained weights, so the memory requirement depends on the number of weights used in the decoder. In our decoder, we reuse the same set of weights $n$ times in each odd layer due to the shift-invariant structure. This effectively reduces the number of weights compared to \cite{Nachmani18}. More precisely, according to the discussion above \eqref{eq:ourodd}, the number of weights in our decoder is $u^2 t$, where $u$ is the number of $1$'s in each row of the parity check matrix, and $t$ is the number of iterations in the BP algorithm. A simple analysis shows that the number of weights in \cite{Nachmani18} is at least $\frac{(n-k)^2}{n} u^2 t$. Therefore, except for extremely high-rate codes where $n-k\le \sqrt{n}$, the memory requirement of our decoder is typically much smaller than \cite{Nachmani18}.

As for the time complexity, we note that the number of additions and multiplications are both proportional to the number of edges in the Tanner graph. The number of edges in the Tanner graph of \cite{Nachmani18} is $u(n-k)$ while for our decoder this number is $un$. Therefore, if we only count the number of additions and multiplications, the ratio between \cite{Nachmani18} and our decoder is $\frac{n-k}{n}$. Although the actual ratio of running time is not exactly $\frac{n-k}{n}$, the above analysis tells us that the running time ratio between \cite{Nachmani18} and our decoder is smaller for high-rate codes and larger for low-rate codes. Table~\ref{tb:time} compares the decoding time of the vanilla versions of \cite{Nachmani18} and our decoder, i.e., no boosting, no list decoding.

In Tables~\ref{tab:ber}--\ref{tab:list} and Fig.~\ref{fig:simu}, we use the $(n-k)\times n$ parity matrix of the form \eqref{eq:generalpc} for the FF neural decoder and hyper graph decoder, and we use the $n\times n$ cyclic parity matrix for our decoder. To evaluate the contribution of the $n\times n$ cyclic matrix, we ran an ablation analysis, where we further consider another two cases: (i) we append $k$ random parity rows to matrix \eqref{eq:generalpc} so that we obtain an  randomly extended $n\times n$ matrix, and we use the FF neural decoder together with this matrix; (ii) we use the FF neural decoder together with the $n\times n$ cyclic matrix. Simulation results in Table~\ref{tab:ablation} show that using randomly extended matrix has very little effect on the BER while using the $n\times n$ cyclic matrix for the FF decoder improves over the $(n-k)\times n$ matrix by $0.45$dB, so what matters here is not the matrix size, but the cyclic structure. When both using the $n\times n$ cyclic matrix, our decoder still demonstrates a $0.25$dB improvement over the FF decoder.

\clearpage

\bibliography{cyclic}
\bibliographystyle{icml2021}


\clearpage

\appendix

\onecolumn

\section{The parity check matrix used in our neural decoder and the permutations used in the list decoding algorithm}

We first describe how to produce the generator matrix and the parity check matrix for BCH codes and punctured RM codes with code length $n=2^m-1$ in our decoding algorithm.\\
{\bf Step 1}: Find a primitive element $\alpha$ of the finite field $\mathbb{F}_{2^m}$. Then the elements of this finite field are $0,1,\alpha,\alpha^2,\dots,\alpha^{2^m-2}$.\\
{\bf Step 2}: Find the generator polynomial of the code. For $1\le j\le 2^m-2$, let $M^{(j)}(x)$ be the minimal polynomial of $\alpha^j$ over the binary field. For BCH code with designed distance $2\delta+1$, the generator polynomial is $g(x)=\lcm\{M^{(1)}(x),M^{(3)}(x),\dots,M^{(2\delta-1)}(x)\}$, where $\lcm$ stands for least common multiple; see Chapter 7.6 of \cite{Macwilliams77}. For $r$th order punctured RM code, the generator polynomial is 
\begin{equation} \label{eq:RMG}
g(x)=\lcm\{M^{(j)}(x): 1\le j\le 2^m-2, ~ 1\le w_2(j)\le m-r-1 \},
\end{equation}
where $w_2(j)$ is the number of $1$'s in the binary expansion of $j$. For example, $w_2(3)=2$ because the binary expansion of $3$ is $(1,1)$, and $w_2(5)=2$ because the binary expansion of $5$ is $(1,0,1)$. See Theorem 11 in Chapter 13.5 of \cite{Macwilliams77} for a proof of \eqref{eq:RMG}.
For an $(n,k)$ cyclic code, the degree of the generator polynomial is $n-k$, so $g(x)$ can be written as $g(x)=g_0+g_1 x+ g_2 x^2 + \dots + g_{n-k} x^{n-k}$, where the coefficients $g_0,g_1,\dots,g_{n-k}$ are either $0$ or $1$. Then the following $k\times n$ matrix is a generator matrix of the code:
\begin{equation} \label{eq:generator}
\begin{array}{ccccccccc}
g_0 & g_1 & g_2 & \dots & g_{n-k} & 0 & 0 &\dots & 0 \\
0 & g_0 & g_1 & g_2 & \dots & g_{n-k} & 0 & \dots & 0 \\
\vdots & \vdots & \vdots & \vdots & \vdots & \vdots & \vdots & \vdots & \vdots \\
0 & 0 & \dots & 0 & g_0 & g_1 & g_2 & \dots & g_{n-k}
\end{array} .
\end{equation}
{\bf Step 3}: For cyclic codes, it always holds that $g(x)$ divides $x^n-1$, and the parity check polynomial of the code is $h(x)=(x^n-1)/g(x)$. Since the degree of $g$ is $n-k$, the degree of $h$ is $k$, and so $h(x)$ can be written as $h(x)=h_k x^k + \dots + h_2 x^2+ h_1 x + h_0$, where the coefficients $h_k,\dots,h_2,h_1,h_0$ are either $0$ or $1$. This explains how to obtain $h_k,\dots,h_2,h_1,h_0$ in the parity matrix \eqref{eq:generalpc}. As already mentioned in Section~\ref{sect:our}, matrix \eqref{eq:generalpc} is used in \cite{Nachmani16,Nachmani18}. In contrast, we use an $n\times n$ parity check matrix consisting of all the $n$ cyclic shifts of the first row of matrix \eqref{eq:generalpc} in our decoder.

Next we describe how to find the permutations used in the list decoding algorithm. More precisely, we explain how to find the set $\mathcal{S}$ consisting of $n+1=2^m$ permutations; see the beginning part of Section~\ref{sect:list}. Note that for all BCH codes and punctured RM codes with code length $n$, we use the {\bf same} set $\mathcal{S}$ of permutations. In other words, the set $\mathcal{S}$ only depends on the code length $n$, and it does not change with code dimension or other parameters.

It is well known that both RM codes and extended BCH codes are invariant to the affine group \cite{Kasami67}. Since $\mathcal{S}$ is a subset of this group, let us begin with describing the affine group. Let $\mathcal{C}$ be a BCH code or a punctured RM code with code length $n$, and let $G$ be its generator matrix obtained from the procedure described above, so $G$ is of the form \eqref{eq:generator}. Let $(C_1,C_2,\dots,C_n)$ be a codeword generated from the matrix $G$. Notice that the matrix $G$ specifies the order of coordinates in the codeword. (For example, swapping two columns of $G$ amounts to swapping the two corresponding coordinates in the codeword.) As already mentioned in Section~\ref{sect:list}, by prepending an overall parity bit $C_0$ we obtain $(C_0,C_1,\dots,C_n)$, a codeword from an extended BCH code or a RM code with length $n+1$. Next we define a one-to-one mapping $f$ between the index set $\{0,1,\dots,n\}$ and the finite field $\mathbb{F}_{2^m}=\{0,1,\alpha,\alpha^2,\dots,\alpha^{n-1}\}$:
$$
f(0)=0, \quad
f(i)=\alpha^{i-1} \text{~~for~} i\in[n].
$$
For $a,b\in\mathbb{F}_{2^m}, a\neq 0$, the affine mapping $X\mapsto aX+b$ defines a permutation on the finite field $\mathbb{F}_{2^m}$, and through the function $f$ it also induces a permutation on the index set $\{0,1,\dots,n\}$. More precisely, for $1\le i\le n$ and $0\le j\le n$, we use $\sigma_{i,j}$ to denote the permutation on $\{0,1,\dots,n\}$ induced by the mapping $X\mapsto f(i)X+f(j)$:
$$
\sigma_{i,j}(v)= f^{-1} \big( f(i) f(v) + f(j) \big) \text{~~for~} v\in\{0,1,\dots,n\}.
$$
The permutations $\{\sigma_{i,j}: 1\le i\le n, 0\le j\le n\}$ form the affine group to which the extended code is invariant.

For the special case of $j=0$, we have
$$
\sigma_{i,0}(0)=0, \quad
\sigma_{i,0}(v)= f^{-1}(\alpha^{i+v-2})
=\left\{ \begin{array}{cc}
  v+i-1   & \text{for~} 1\le v\le n-i+1 \\
  v+i-1-n   & \text{for~} n-i+2 \le v\le n
\end{array} \right. .
$$
Therefore, $\sigma_{i,0}$ is the permutation that fixes $C_0$ and performs $(i-1)$ cyclic right shifts on $(C_1,C_2,\dots,C_n)$. It is not a surprise that the extended code is invariant to such a permutation because $(C_1,C_2,\dots,C_n)$ belongs to a cyclic code.

For the purpose of list decoding, we focus on another special case $i=1$, and we write $\sigma_j=\sigma_{1,j}$ to simplify the notation. By definition, $\sigma_j$ is the permutation on $\{0,1,\dots,n\}$ induced by the mapping $X\mapsto X+f(j)$, so $\sigma_j(v)=f^{-1}(f(v)+f(j))$ for $0\le v\le n$. Clearly, $\sigma_0$ is the identity permutation. The set $\mathcal{S}$ used in our list decoding algorithm is $\mathcal{S}=\{\sigma_0,\sigma_1,\dots,\sigma_n\}$. Here we give a concrete example for $n=15$. Each row in the following matrix represents a permutation $\sigma_j$. From top to bottom, these permutations are $\sigma_0,\sigma_1,\sigma_2,\sigma_5,\sigma_3,\sigma_9,\sigma_6,\sigma_{11},\sigma_4,\sigma_{15},\sigma_{10},\sigma_8,\sigma_7,\sigma_{14},\sigma_{12},\sigma_{13}$.
$$
\begin{array}{cccccccccccccccc}
0 & 1 & 2 & 3 & 4 & 5 & 6 & 7 & 8 & 9 & 10 & 11 & 12 & 13 & 14 & 15 \\
1 & 0 & 5 & 9 & 15 & 2 & 11 & 14 & 10 & 3 & 8 & 6 & 13 & 12 & 7 & 4 \\
2 & 5 & 0 & 6 & 10 & 1 & 3 & 12 & 15 & 11 & 4 & 9 & 7 & 14 & 13 & 8 \\
5 & 2 & 1 & 11 & 8 & 0 & 9 & 13 & 4 & 6 & 15 & 3 & 14 & 7 & 12 & 10 \\
3 & 9 & 6 & 0 & 7 & 11 & 2 & 4 & 13 & 1 & 12 & 5 & 10 & 8 & 15 & 14 \\
9 & 3 & 11 & 1 & 14 & 6 & 5 & 15 & 12 & 0 & 13 & 2 & 8 & 10 & 4 & 7 \\
6 & 11 & 3 & 2 & 12 & 9 & 0 & 10 & 14 & 5 & 7 & 1 & 4 & 15 & 8 & 13 \\
11 & 6 & 9 & 5 & 13 & 3 & 1 & 8 & 7 & 2 & 14 & 0 & 15 & 4 & 10 & 12 \\
4 & 15 & 10 & 7 & 0 & 8 & 12 & 3 & 5 & 14 & 2 & 13 & 6 & 11 & 9 & 1 \\
15 & 4 & 8 & 14 & 1 & 10 & 13 & 9 & 2 & 7 & 5 & 12 & 11 & 6 & 3 & 0 \\
10 & 8 & 4 & 12 & 2 & 15 & 7 & 6 & 1 & 13 & 0 & 14 & 3 & 9 & 11 & 5 \\
8 & 10 & 15 & 13 & 5 & 4 & 14 & 11 & 0 & 12 & 1 & 7 & 9 & 3 & 6 & 2 \\
7 & 14 & 12 & 4 & 3 & 13 & 10 & 0 & 11 & 15 & 6 & 8 & 2 & 5 & 1 & 9 \\
14 & 7 & 13 & 15 & 9 & 12 & 8 & 1 & 6 & 4 & 11 & 10 & 5 & 2 & 0 & 3 \\
12 & 13 & 7 & 10 & 6 & 14 & 4 & 2 & 9 & 8 & 3 & 15 & 0 & 1 & 5 & 11 \\
13 & 12 & 14 & 8 & 11 & 7 & 15 & 5 & 3 & 10 & 9 & 4 & 1 & 0 & 2 & 6
\end{array}
$$
As a final remark, all the methods described above can be efficiently programmed using the Communications Toolbox of Matlab. The Matlab code is available at \href{https://github.com/cyclicallyneuraldecoder/CyclicallyEquivariantNeuralDecoders}{github.com/cyclicallyneuraldecoder}

\section{Our neural decoder is equivariant to all cyclic shifts}
Recall the definition of cyclic shifts $\pi_j, j\in[n]$ in \eqref{eq:permu}. Observe that $(C_{\pi_j(1)}, C_{\pi_j(2)}, \dots, C_{\pi_j(n)})$ is obtained by $(j-1)$ cyclic left shifts of $(C_1,C_2,\dots,C_n)$. In particular, $(C_{\pi_2(1)}, C_{\pi_2(2)}, \dots, C_{\pi_2(n)})=(C_2,C_3,\dots,C_n,C_1)$ is obtained by one cyclic left shift of $(C_1,C_2,\dots,C_n)$.

Let $(L_1,L_2,\dots,L_n)$ be an LLR vector and let $(o_1,o_2,\dots,o_n)$ be the corresponding decoding result of our neural decoder. We will prove that for every $j\in[n]$, if the LLR vector is $(L_{\pi_j(1)}, L_{\pi_j(2)}, \dots, L_{\pi_j(n)})$, then the decoding result of our decoder becomes $(o_{\pi_j(1)}, o_{\pi_j(2)}, \dots, o_{\pi_j(n)})$. In fact, we only need to prove this claim for $j=2$, and the claim for other values of $j$ follows by a simple induction. Below we will write $\pi=\pi_2$ to simplify the notation.

Recall that we use \eqref{eq:classiceven} to calculate the messages in even layers and we use \eqref{eq:ourodd} for odd layers. Finally, we use \eqref{eq:ourout} to calculate the output layer. Also recall that $E$ is the set of edges in the Tanner graph. Let $x^{[s]}(e), e\in E$ be the messages in the $s$th layer when the input LLR vector is $(L_1,L_2,\dots,L_n)$, and let $\tilde{x}^{[s]}(e), e\in E$ be the messages in the $s$th layer when the input LLR vector is $(\tilde{L}_1,\tilde{L}_2,\dots,\tilde{L}_n)=(L_{\pi(1)}, L_{\pi(2)}, \dots, L_{\pi(n)})$. We will prove that 
\begin{equation} \label{eq:bc}
\tilde{x}^{[s]}((c_i,v_j))=x^{[s]}((c_{\pi(i)},v_{\pi(j)})) \text{~~for all~} (c_i,v_j)\in E.
\end{equation}
Notice that whenever $(c_i,v_j)\in E$, we always have $(c_{\pi(i)},v_{\pi(j)})\in E$; see Fig.~\ref{fig:our_matrix} for an illustration.

We prove \eqref{eq:bc} by induction on $s$. It holds for $s=0$ because we set the initialization (both $x^{[0]}$ and $\tilde{x}^{[0]}$) to be the all-zero vector in our algorithm. This establishes the induction base.

Now let $s$ be an odd number and suppose that \eqref{eq:bc} holds for $s-1$. Let us prove that \eqref{eq:bc} also holds for $s$. By \eqref{eq:ourodd},
\begin{align*}
\tilde{x}^{[s]}  ((c_{\pi_j(i_b)}, v_j))  
& = \tanh \Big( \frac{1}{2} \Big( w_b^{[s]} \tilde{L}_j
 + \sum_{b'\in [u]\setminus \{b\} }
w_{b', b}^{[s]} ~
\tilde{x}^{[s-1]}((c_{\pi_j(i_{b'})}, v_j)) \Big) \Big) \\
& = \tanh \Big( \frac{1}{2} \Big( w_b^{[s]} L_{\pi(j)}
 + \sum_{b'\in [u]\setminus \{b\} }
w_{b', b}^{[s]} ~
x^{[s-1]}((c_{\pi(\pi_j(i_{b'}))}, v_{\pi(j)})) \Big) \Big) \\
& = x^{[s]}((c_{\pi(\pi_j(i_b))}, v_{\pi(j)})) ,
\end{align*}
where the second equality follows from the induction hypothesis.

Now let $s$ be an even number and suppose that \eqref{eq:bc} holds for $s-1$. Let us prove that \eqref{eq:bc} also holds for $s$. By \eqref{eq:classiceven},
\begin{align*}
\tilde{x}^{[s]}((c_i,v_j)) 
& = 2 \tanh^{-1} \Big( \prod_{(c_i,v_{j'})\in N(c_i)\setminus \{(c_i,v_j)\} } \tilde{x}^{[s-1]}((c_i,v_{j'})) \Big) \\
& = 2 \tanh^{-1} \Big( \prod_{(c_i,v_{j'})\in N(c_i)\setminus \{(c_i,v_j)\} } x^{[s-1]}((c_{\pi(i)},v_{\pi(j')})) \Big) \\
& = 2 \tanh^{-1} \Big( \prod_{(c_{\pi(i)},v_{\pi(j')})\in N(c_{\pi(i)})\setminus \{(c_{\pi(i)},v_{\pi(j)})\} } x^{[s-1]}((c_{\pi(i)},v_{\pi(j')})) \Big) \\
& =x^{[s]}((c_{\pi(i)},v_{\pi(j)})) .
\end{align*}
This establishes the inductive step and completes the proof of \eqref{eq:bc}.

Finally, denote the output corresponding to $(\tilde{L}_1,\tilde{L}_2,\dots,\tilde{L}_n)$ as $(\tilde{o}_1,\tilde{o}_2,\dots,\tilde{o}_n)$. Then by \eqref{eq:ourout},
\begin{align*}
\tilde{o}_j & = \tilde{L}_j  +  \sum_{b\in[u]} 
w_b^{\out} ~
\tilde{x}^{[2t]}((c_{\pi_j(i_b)}, v_j)) \\
& = L_{\pi(j)}  +  \sum_{b\in[u]} 
w_b^{\out} ~
x^{[2t]}((c_{\pi(\pi_j(i_b))}, v_{\pi(j)})) \\
& = o_{\pi(j)} .
\end{align*}
Thus we have proved that our neural decoder is equivariant to all cyclic shifts.

\section{More plots of the simulation results}

\begin{figure*}
\centering
\begin{subfigure}{0.49\textwidth}
\centering
\includegraphics[width=\textwidth]{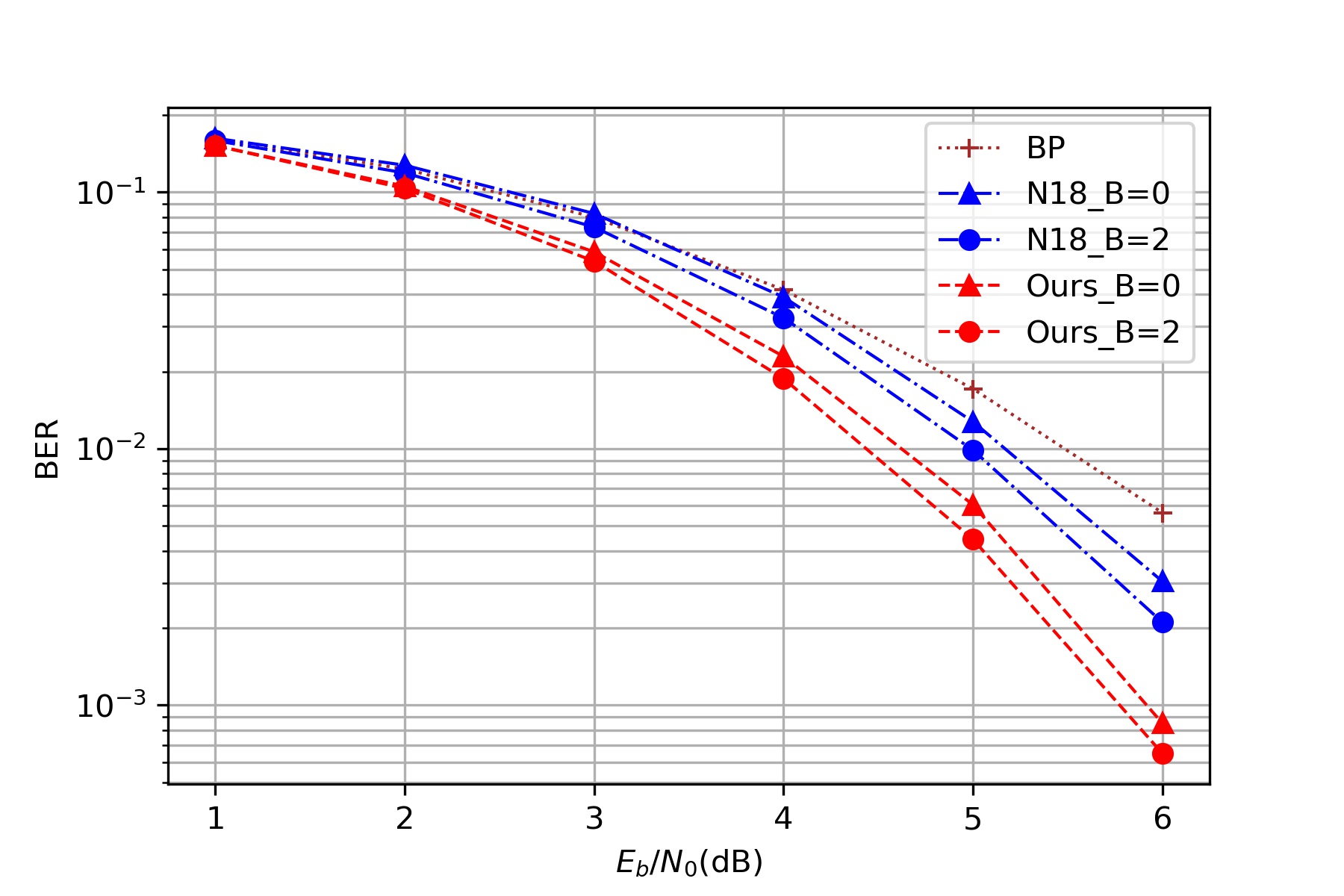} 
\caption{BCH(63,24)}
\end{subfigure}
~
\begin{subfigure}{0.49\textwidth}
\centering
\includegraphics[width=\textwidth]{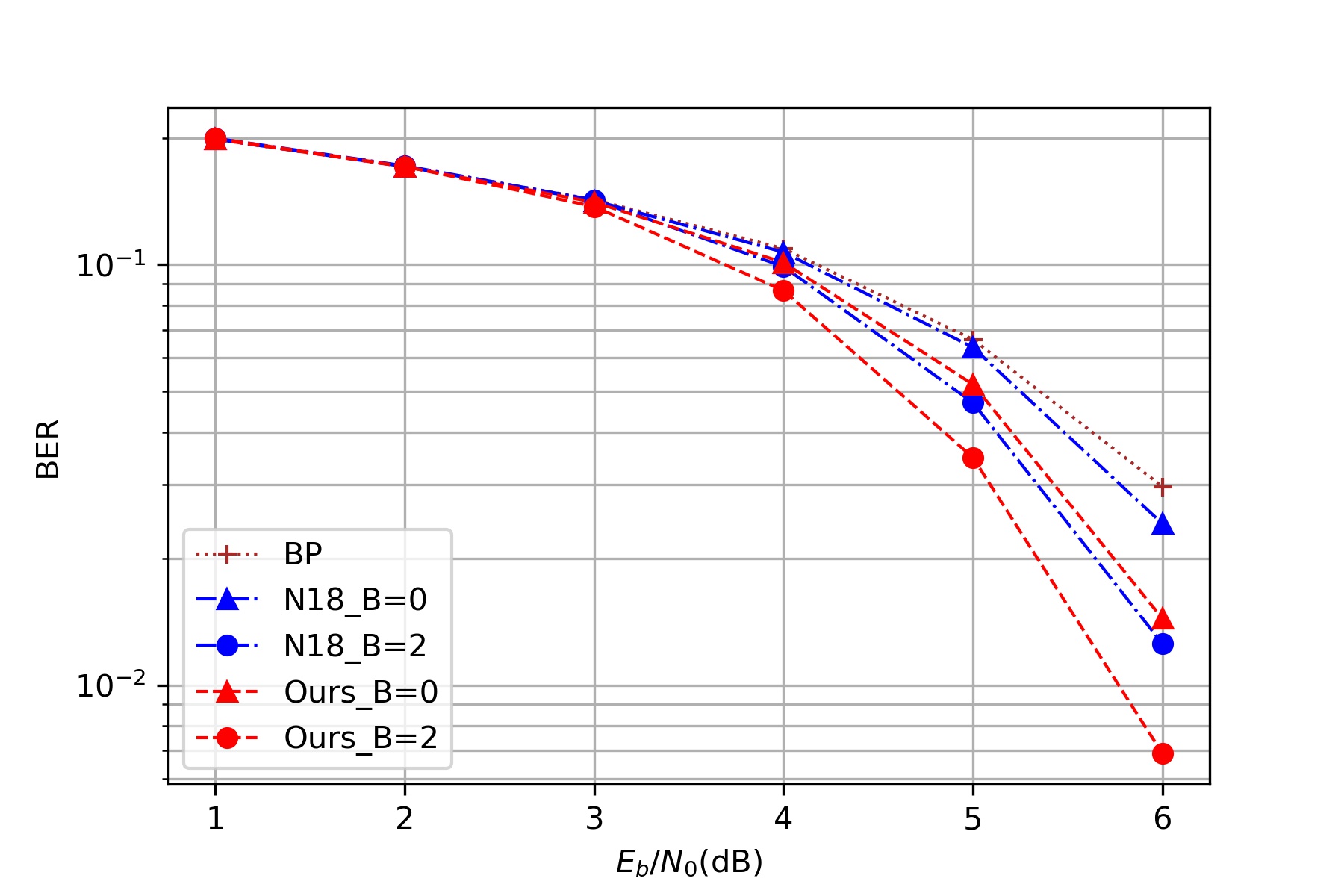} 
\caption{BCH(127,36)}
\end{subfigure}

\begin{subfigure}{0.49\textwidth}
\centering
\includegraphics[width=\textwidth]{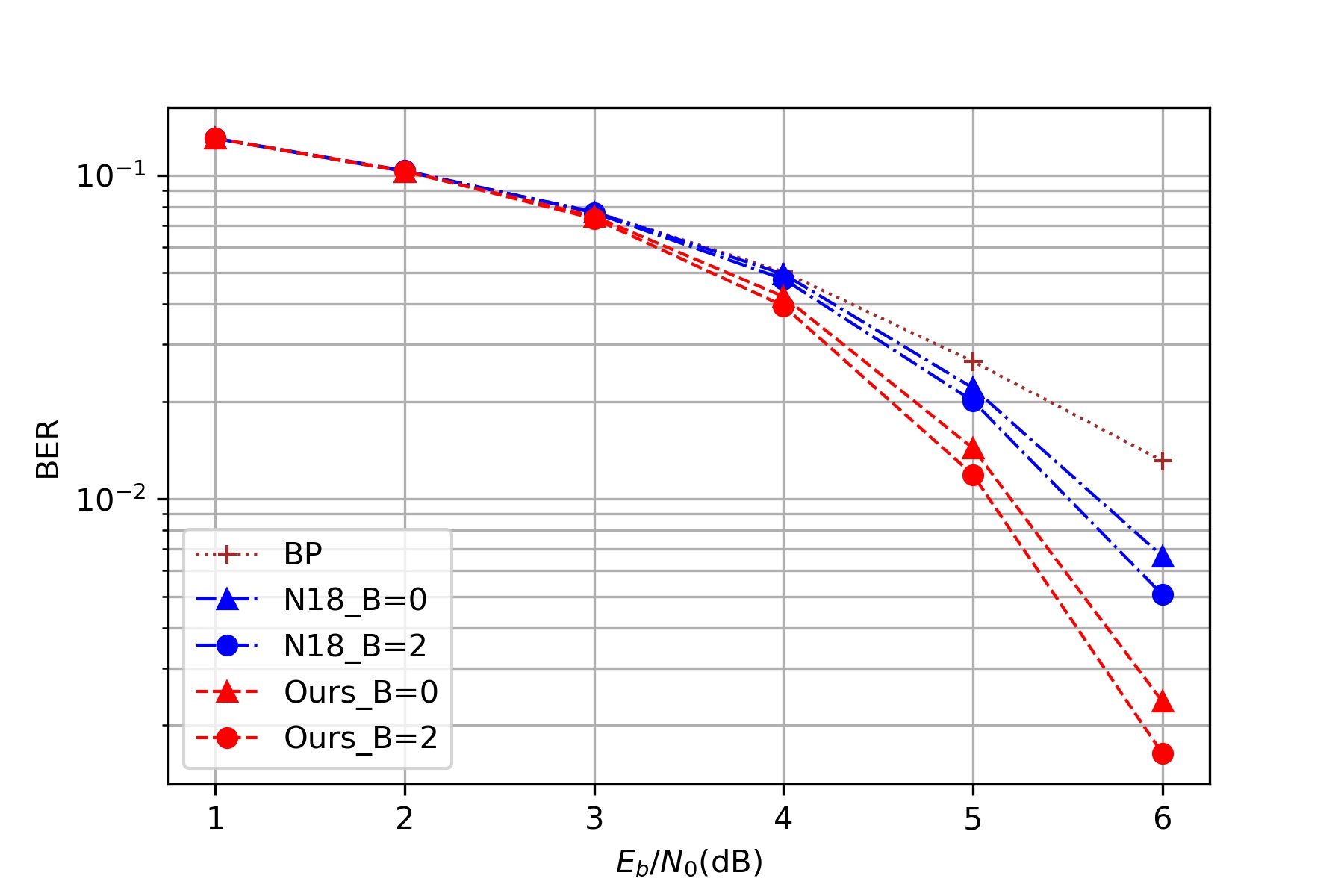} 
\caption{BCH(127,64)}
\end{subfigure}
~
\begin{subfigure}{0.49\textwidth}
\centering
\includegraphics[width=\textwidth]{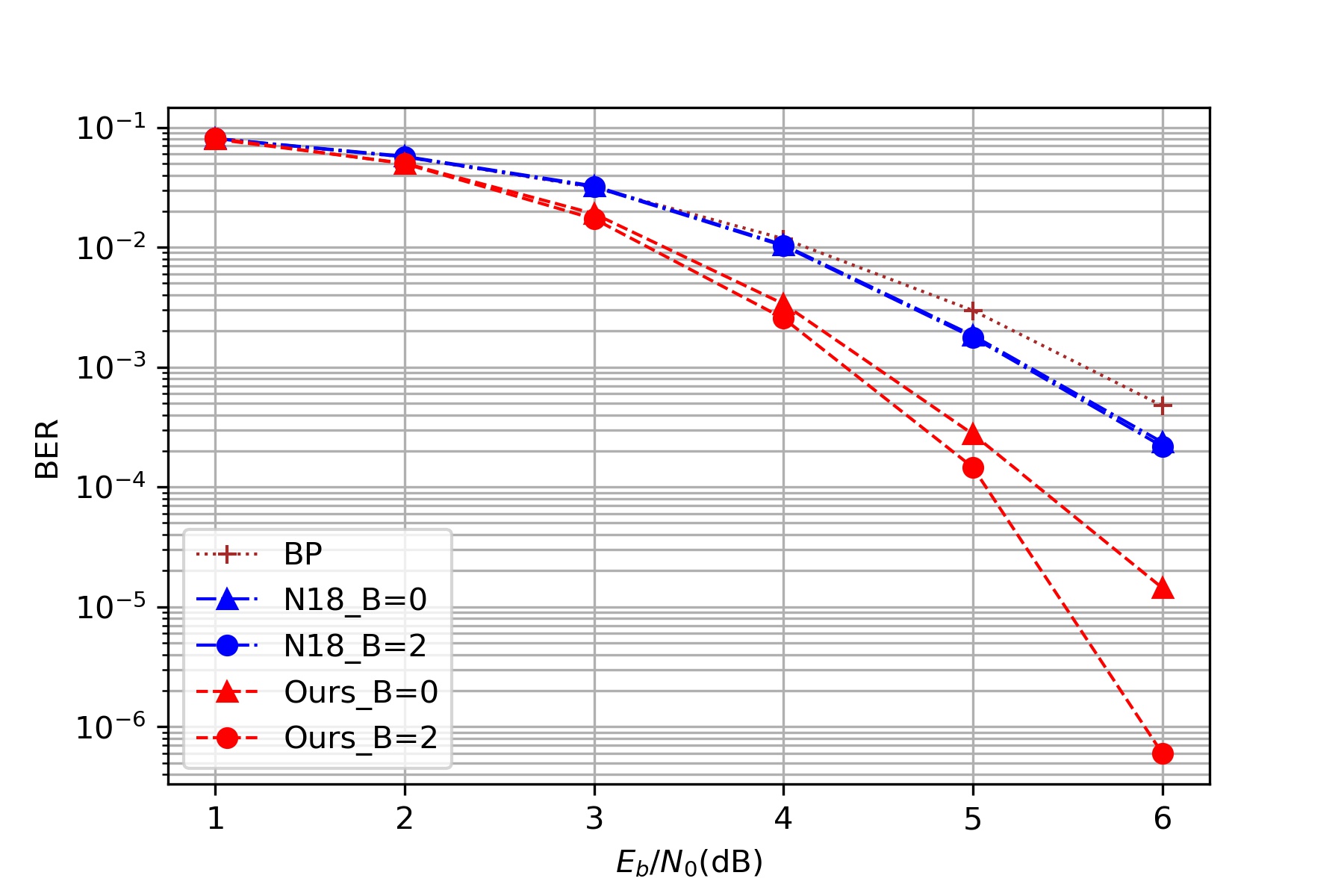} 
\caption{Punctured RM(127,99)}
\end{subfigure}

\begin{subfigure}{0.49\textwidth}
\centering
\includegraphics[width=\textwidth]{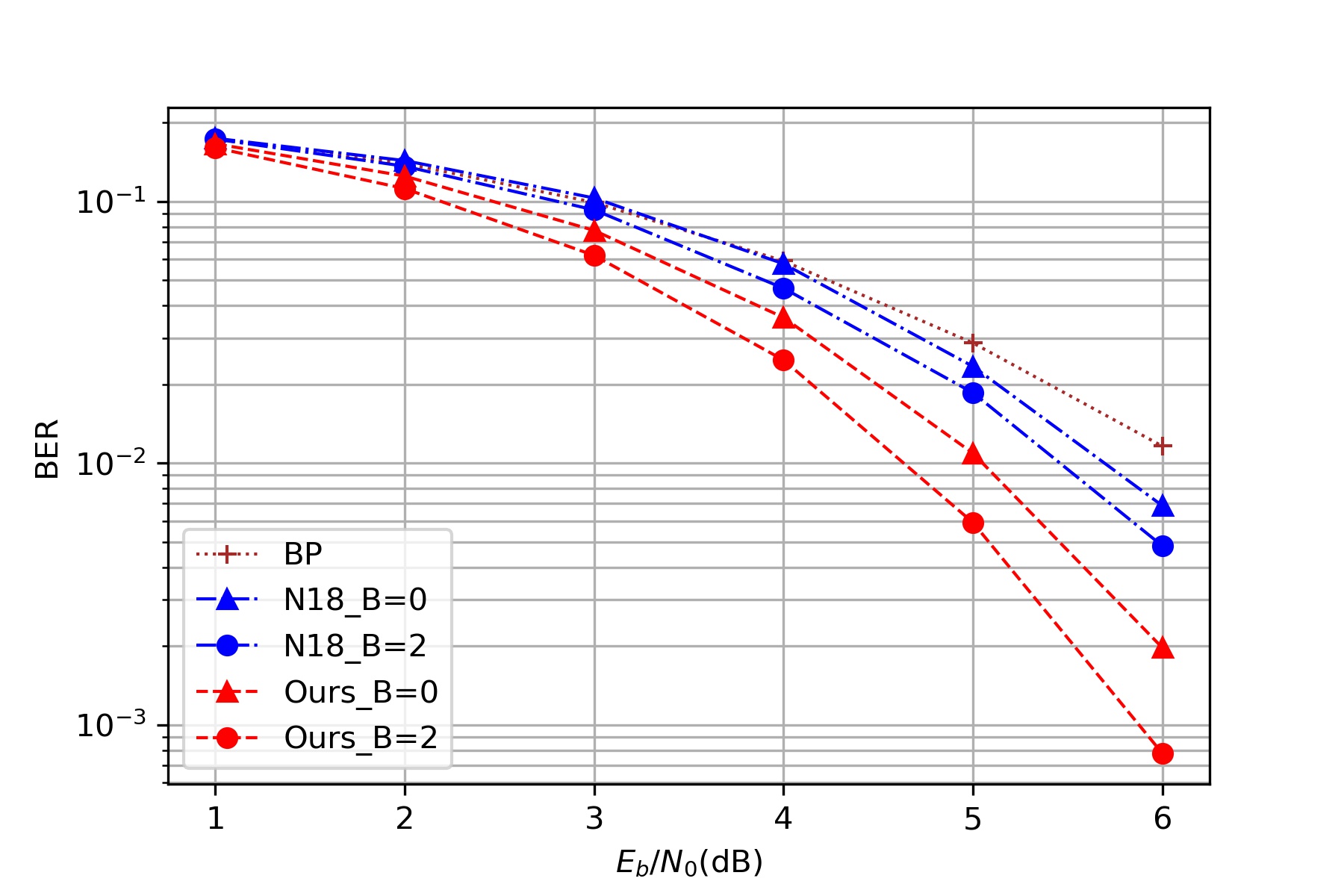} 
\caption{Punctured RM(63,22)}
\end{subfigure}
~
\begin{subfigure}{0.49\textwidth}
\centering
\includegraphics[width=\textwidth]{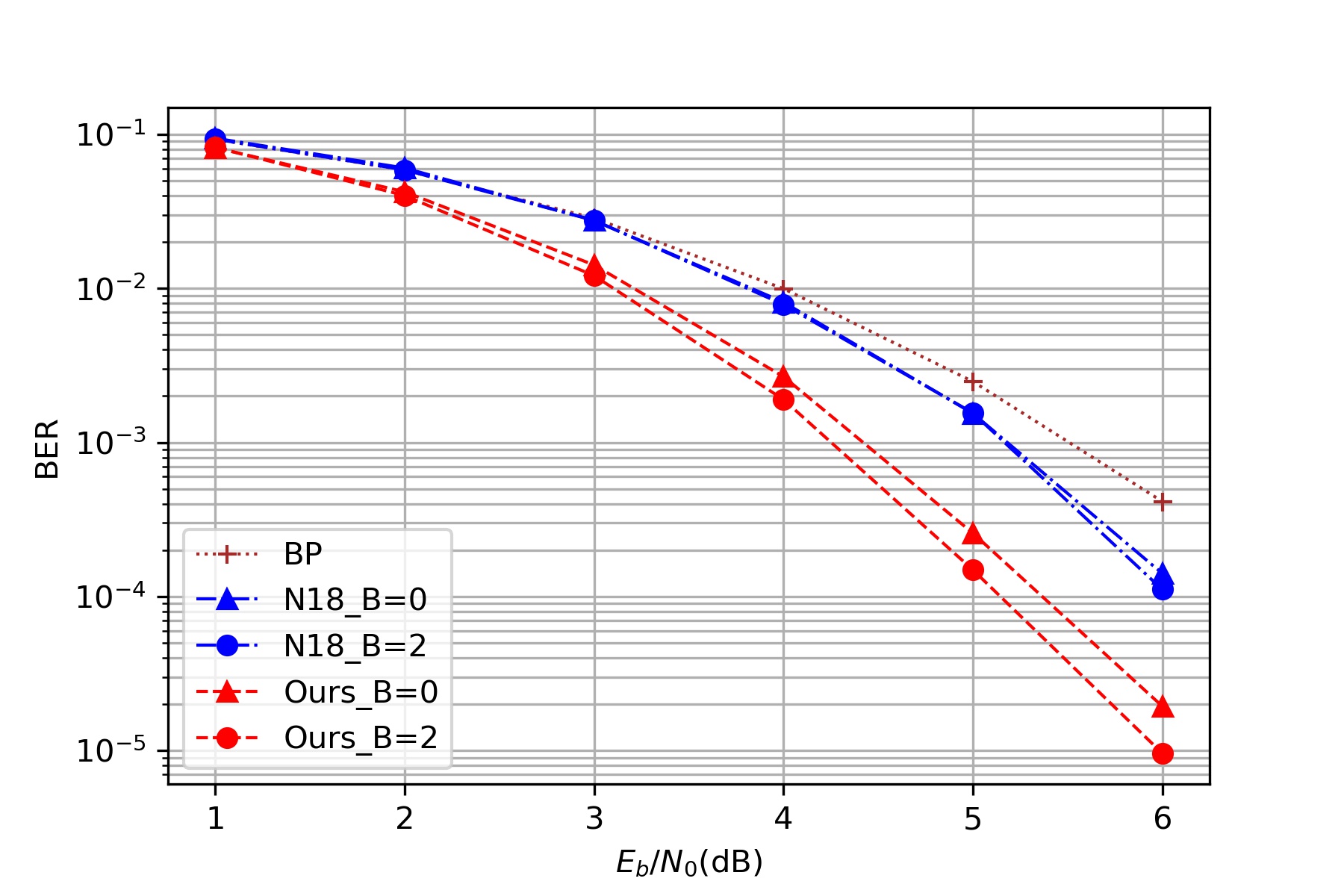} 
\caption{Punctured RM(63,42)}
\end{subfigure}
\end{figure*}

\begin{figure*}
\begin{subfigure}{0.49\textwidth}
\centering
\includegraphics[width=\textwidth]{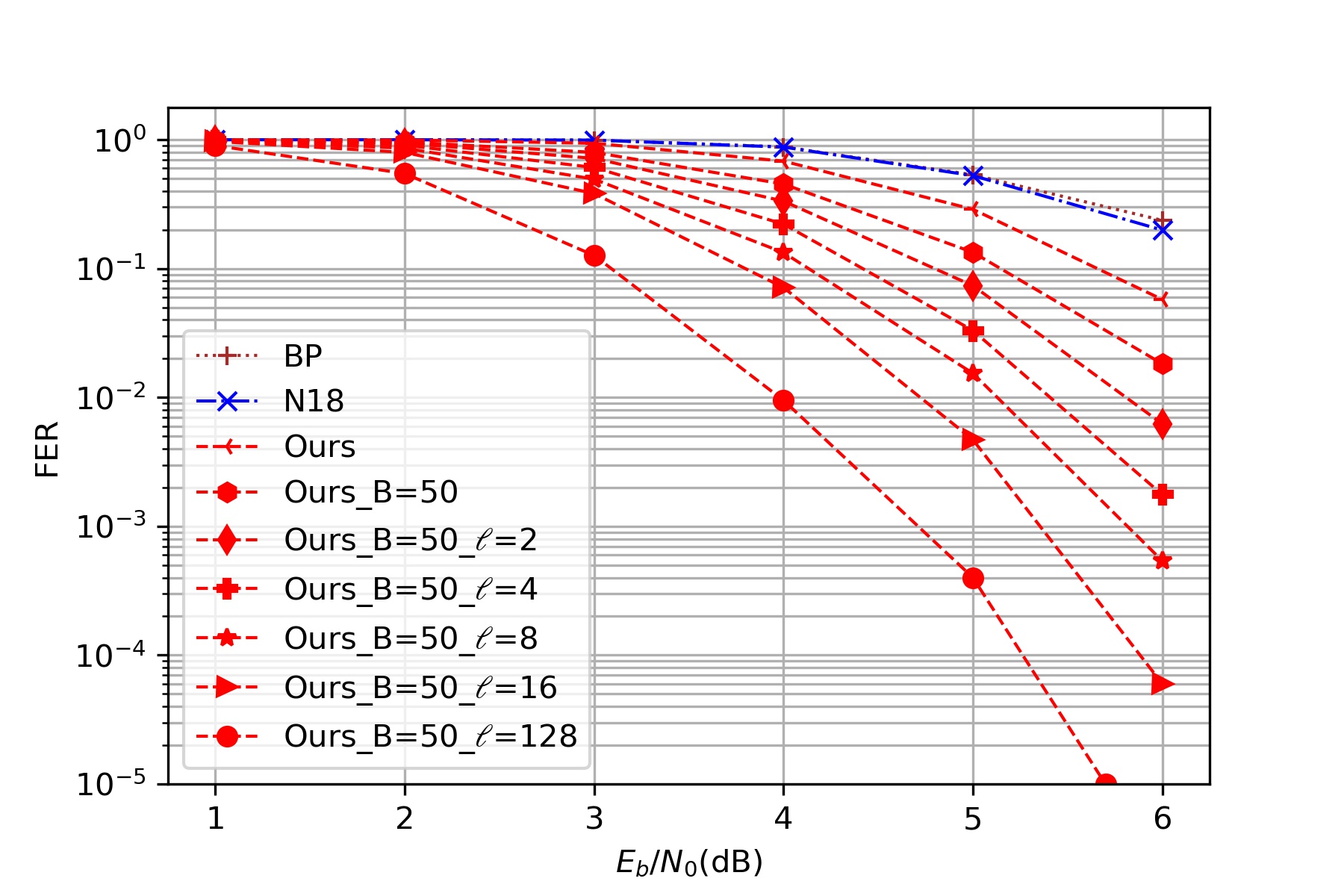} 
\caption{BCH(127,64) List decoding}
\end{subfigure}
~
\begin{subfigure}{0.49\textwidth}
\centering
\includegraphics[width=\textwidth]{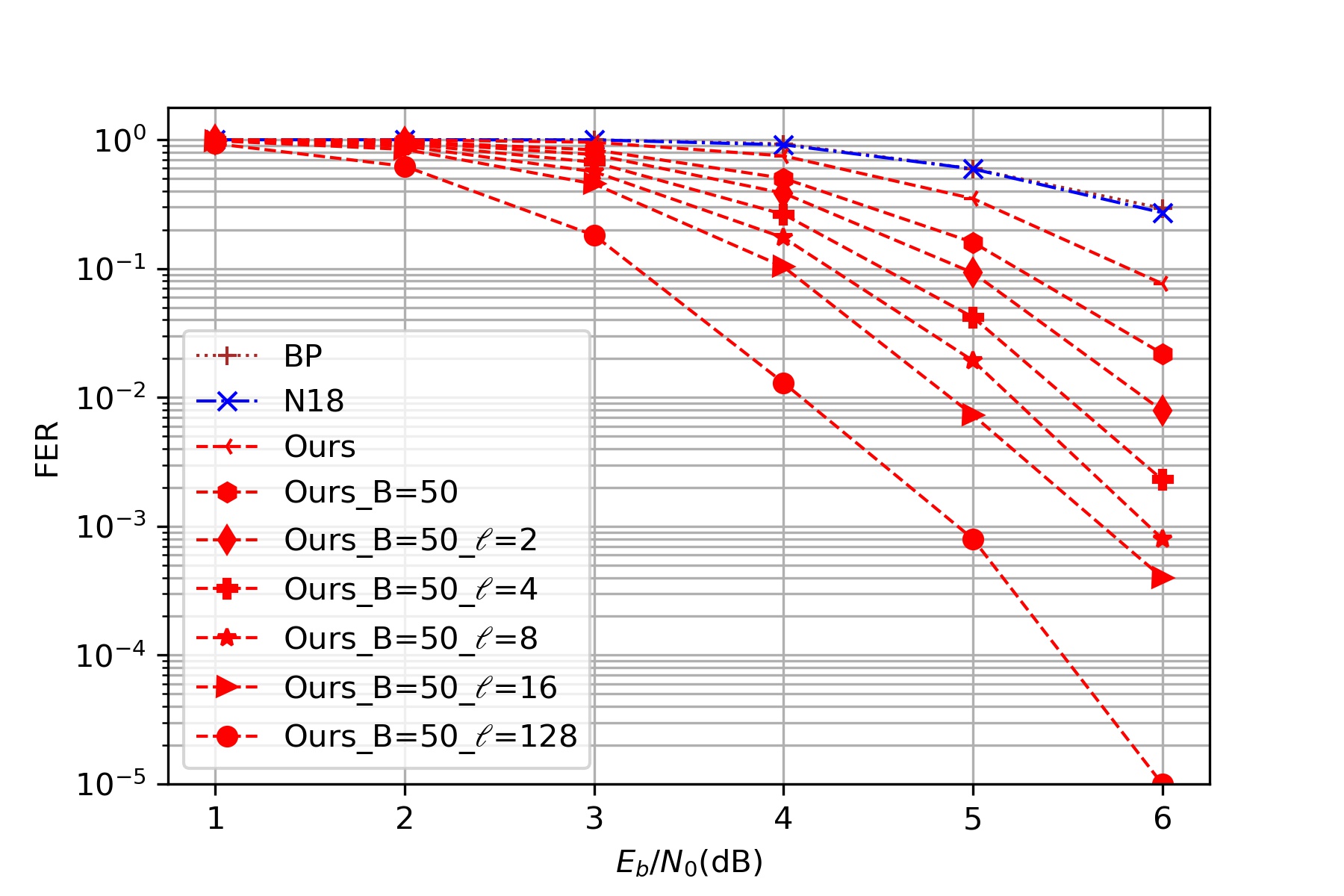} 
\caption{Punctured RM(127,64) List decoding}
\end{subfigure}
\end{figure*}


\end{document}